%% file: main.tex
\documentclass[%
reprint,
amsmath,
amssymb,
aps,
pra,
floatfix,
nofootinbib
]{revtex4-1}

\input{preamble}

\begin{document}

\title{Automated discovery of heralded ballistic graph state generators for fusion-based photonic quantum computation}

\author{Gavin~S.~Hartnett}
\email{gavin.hartnett@q-ctrl.com}
\author{Dave~Kielpinski}
\altaffiliation{Current address: Nearmap, Level 4 Tower One, International Towers, Barangaroo NSW 2000.}
\author{Smarak~Maity}
\author{Pranav~S.~Mundada}
\author{Yuval~Baum}
\author{Michael~R.~Hush}
\affiliation{Q-CTRL, Los Angeles, CA USA and Sydney, NSW Australia}

\begin{abstract} 
Designing photonic circuits that prepare graph states with high fidelity and success probability is a central challenge in linear optical quantum computing. 
Existing approaches rely on hand-crafted designs or fusion-based assemblies. 
In the absence of multiplexing/boosting, both post-selected ballistic circuits and sequential fusion exhibit exponentially decreasing single-shot yields - a fundamental limitation that makes optimizing individual resource state generators particularly important, as these serve as building blocks in larger FBQC architectures.
We present a general-purpose optimization framework for automated photonic circuit discovery using a novel polynomial-based simulation approach, enabling efficient strong simulation and gradient-based optimization. Our framework employs a two-pass optimization procedure: the first pass identifies a unitary transformation that prepares the desired state with perfect fidelity and maximal success probability, and the second pass implements a novel sparsification algorithm that reduces this solution to a compact photonic circuit with minimal beamsplitter count while preserving performance. This sparsification procedure often reveals underlying mathematical structure, producing highly simplified circuits with rational reflection coefficients. We demonstrate our approach by discovering optimized circuits for $3$-, $4$-, and $5$-qubit graph states across multiple equivalence classes. For 4-qubit states, our circuits achieve success probabilities of $2.053 \times 10^{-3}$ to $7.813 \times 10^{-3}$, outperforming the fusion baseline by up to $4.7 \times$. For 5-qubit states, we achieve $5.926 \times 10^{-5}$ to $1.157 \times 10^{-3}$, demonstrating up to $7.5 \times$ improvement. These results include the first known state preparation circuits for certain 5-qubit graph states.
\end{abstract}

\maketitle

\section{Introduction}

The seminal Knill-Laflamme-Milburn (KLM) protocol established that universal quantum computation is achievable using only single-photon sources, linear optics, and adaptive photon detection~\cite{knill_scheme_2001}.
Building on this foundation, a sequence of architecture-level advances---ranging from optical cluster-state proposals~\cite{nielsen_optical_2004} to more recent schemes with built-in fault tolerance~\cite{bartolucci_fusion-based_2023}---has markedly reduced resource overheads. In discrete-variable (dual-rail) photonic approaches, a common requirement is the high-fidelity, high-rate generation of graph states that serve as resources for universal Measurement-Based Quantum Computing (MBQC)~\cite{raussendorf_one-way_2001, raussendorf_measurement-based_2003}, either as large clusters or as a steady supply of small resource states for fusion. However, current on-chip demonstrations remain at roughly the 8-12-photon scale, and free-space experiments, while sometimes larger, operate at low rates~\cite{zhong_12-photon_2018, lee_quantum_2024}---still far below what is needed for even simple fault-tolerant algorithms.

Fusion-Based Quantum Computing (FBQC) addresses this gap by stitching together many small entangled resource states using probabilistic, heralded two-photon Bell measurements (fusion operations)~\cite{browne_resource-efficient_2005, bartolucci_fusion-based_2023}. Although a single-shot, unboosted analysis suggests an exponential suppression in success as target size grows, two standard remedies restore practicality: (i) percolation schemes that accept a random cluster once the bond success exceeds the relevant lattice percolation threshold~\cite{gimeno-segovia_three-photon_2015, kieling_percolation_2007, pant_percolation_2019}; and (ii) repeat-until-success (RUS) and other boosted fusion variants that raise success using ancilla photons and switching (often with multiplexed sources)~\cite{lim_repeat-until-success_2005}. In practice, FBQC leverages the latter together with fault-tolerant decoding on a fixed fusion network, so it does not require percolation. Despite recent demonstrations of boosted fusion rates above typical percolation thresholds~\cite{guo_boosted_2024}, experimental demonstrations of graph states assembled by such fusions remain modest in size and rate (often $\lesssim 10$ photons on chip~\cite{lu_experimental_2007}), still far from fault-tolerant scales.

In linear-optical FBQC, overall performance is driven by several factors, with a central lever being the availability of high-fidelity, high-throughput heralded resource-state generators (RSGs). Finding RSGs that maximize success probability while minimizing optical complexity under loss remains a challenging open problem. We introduce a general-purpose optimization framework and carry out a systematic search up to five qubits, reporting numerically discovered RSGs for all 3- and 4-qubit graph states and a broad subset of 5-qubit graphs; to our knowledge, this is the most comprehensive automated search over heralded, passive, dual-rail photonic circuits for 3-5-qubit graph states to date.

\paragraph*{Current state of the art.}
Work on RSGs has concentrated on dual-rail GHZ states, which can subsequently be converted or fused into more general graph states \cite{gimeno-segovia_three-photon_2015}.
The analytical framework reported by Bartolucci {\em et al.} currently attains the highest published success probabilities for $n$-qubit dual-rail GHZ states using fully ballistic linear-optical circuits~\cite{bartolucci_creation_2021}.
Notably, these circuits were constructed manually, leveraging symmetries and tailored circuit identities.
Our own numerical survey corroborates the optimality of that success probability frontier, albeit revealing that multiple inequivalent circuits attain the same bound. Unlike prior handcrafted constructions, our method is automated, enables tractable optimization for systems up to 5 qubits, and is tunable for different hardware constraints---making it better suited to practical deployment in FBQC-based architectures.

\paragraph*{Prior computational investigations of dual-rail heralded ballistic RSGs.} 
The development of automated discovery methods began with numerical searches restricted to small systems.  
Early analytic work and gradient-based numerical search demonstrated that heralded dual-rail Bell states require at least four photons and provided strong numerical evidence for the optimality of known four-photon circuits~\cite{stanisic_generating_2017}.  
Further gradient-based numerical optimization over arbitrary $m$-mode interferometers subsequently pushed the postselected success probabilities for the Bell pair and the 3-qubit GHZ state to 2/27 and 1/54, respectively~\cite{gubarev_improved_2020}.
A follow-up study re-optimized directly in Fock space and post-filtered for implementable linear optics, but remained limited to $\le 6$ modes owing to the exponential scaling of the Fock representation~\cite{gubarev_fock_2021}.  
Genetic-algorithm search has also been applied to heralded 2-qubit gates, achieving results with up to eight modes and four photons~\cite{chernikov_heralded_2023}.  
All of the above investigations are fully ballistic (no adaptive switching) and employ dual-rail encoding.

We note two influential, but non-ballistic numerical studies that use quantum non-demolition (QND) measurements to grow \(n\)-qubit linear cluster states, achieving per-instance success scaling as \((1/2)^{n-1}\) under their assumptions~\cite{uskov_resource-efficient_2015,uskov_optimal_2015}. Because these schemes rely on non-destructive/QND measurements with adaptive feedforward, rather than the passive, destructive, postselected dual-rail resource model we adopt, we exclude them from our comparison.

Lastly, general-purpose optical-design engines such as \textsc{MELVIN}, \textsc{THESEUS}, and \textsc{PyTheus} (Refs.~\cite{krenn_automated_2016, krenn_conceptual_2021, ruiz-gonzalez_digital_2023}) explore vast experiment spaces via stochastic or graph-theoretic heuristics. While \textsc{PyTheus} incorporates automatic differentiation for gradient-based optimization, \textsc{MELVIN} and \textsc{THESEUS} rely on discrete search methods (stochastic and graph-theoretic, respectively) that are not directly compatible with gradient-based optimization. More importantly, none of these tools are specifically optimized for high-probability generation of dual-rail graph states suited to MBQC, and their computational approaches differ fundamentally from the polynomial-based strong simulation framework needed to efficiently scale beyond small photon numbers for heralded ballistic state preparation.

\paragraph*{Efficient automated RSG discovery pipeline} 
In this paper, we present an automated pipeline that is able to discover RSGs for arbitrary target states. We develop a fully differentiable, polynomial-based simulation and optimization engine that autonomously searches the space of linear optical circuits (parameterized as $m$-mode unitary transfer matrices) for heralded state preparation. Using up to 20 spatial modes and 12 photons, the algorithm discovers inequivalent graph state circuits of up to five qubits, delivers success probability gains of up to $4.7\times$ (4-qubit) and $7.5\times$ (5-qubit) over fusion-based baselines, and uncovers previously unknown solutions for several graph topologies.

These achievements required two critical innovations: (1) A differentiable polynomial-based algorithm for the strong linear optical simulation of photonic states. We represent quantum states as polynomials over mode creation operators, enabling an efficient FFT-based convolution algorithm to implement the required polynomial multiplications. To enhance scalability, our algorithm can be configured to simulate only the experimentally-relevant subspace where signal and ancilla modes receive specific photon numbers. Crucially, our implementation is compatible with automatic differentiation, facilitating gradient computation during optimization. (2) A novel circuit sparsification procedure that reduces dense transfer matrices to compact photonic circuits via regularized optimization. This approach penalizes circuit complexity while preserving heralded state fidelities and success probabilities, often revealing underlying mathematical structure that produces significantly sparsified circuits with rational reflection coefficients. The sparsified circuits can be compiled into experimentally feasible implementations using non-local beamsplitters.

Beyond its immediate FBQC impact, our framework demonstrates how gradient-based automated discovery can systematically bridge the gap between abstract photonic architectures and hardware-feasible implementations. In the following sections, we detail both the theoretical foundations and practical achievements of this approach. The remainder of the paper is organized as follows: Section~\ref{sec:background} establishes the mathematical framework for linear optical quantum computing and heralded state preparation. Section~\ref{sec:optimization} presents our two-stage optimization pipeline, beginning with our novel polynomial-based simulation algorithm (Sec.~\ref{sec:simulation}), followed by transfer matrix optimization (Sec.~\ref{sec:optimizationpass1}) and circuit sparsification (Sec.~\ref{sec:optimizationpass2}). Section~\ref{sec:graphstateresults} demonstrates the pipeline's capabilities through systematic application to 3-, 4-, and 5-qubit graph states, with performance comparisons to fusion-based approaches. Section~\ref{sec:discussion} discusses implications and future directions, while Section~\ref{sec:conclusion} concludes.

\section{Background \label{sec:background}}

In this work, we consider photonic quantum circuits with $m$ modes and $n$ photons. The general state may be written as a superposition of Fock states:
\begin{equation}
    \label{eq:general_state}
    \kket{\Psi} = \sum_{(n_1, n_2, ..., n_m) \in \Phi_{m,n}} c_{n_1, n_2, ..., n_m} \kket{n_1, n_2, ..., n_m} \,.
\end{equation}
Here, $n_i \in \mathbb{Z}_+$ represents the $i$-th mode occupation number. The sum is over $\Phi_{m,n}$, the set of $m$-tuples $(n_1, ..., n_m)$ satisfying $\sum_{i=1}^m n_i = n$, corresponding to states with exactly $n$ total photons across all modes. Here, following \cite{bartolucci_creation_2021}, we use the dual-rail notation $\kket{\cdot}$ to denote photonic states, as opposed to logical qubit states. In this notation, the state $\kket{10}$ corresponds to the qubit state $\ket{0}$, and $\kket{01}$ to $\ket{1}$. For convenience, we also use a shorthand where tuples are indicated by a single bold symbol, i.e., we denote $(n_1, n_2, ..., n_m)$ by $\bm{n}$, so that the above may be more compactly written as $\kket{\Psi} = \sum_{\bm{n} \in \Phi_{m,n}} c_{\bm{n}} \kket{\bm{n}}$. Furthermore, each Fock state is generated from the vacuum through the action of creation operators:
\begin{equation}
    \kket{\bm{n}} = \frac{(a_{1}^{\dagger})^{n_1} \ldots (a_{m}^{\dagger})^{n_m}}{\sqrt{n_1! \ldots n_m!}} \kket{\Omega} \,,
\end{equation}
where $\kket{\Omega}$ is the 0-particle vacuum, the creation operator $a_i^{\dagger}$ adds a photon to mode $i$, and the denominator is the usual normalization constant for Fock states. 

Starting from an initial Fock state, the system then transforms under the action of a linear photonic circuit, representable as an $m \times m$ unitary transfer matrix $U$ which evolves the input photon creation operators to output creation operators according to the rule:
\begin{equation}
    \label{eq:LOQCtransformation}
    a^{\dagger}_{\text{in}, i} = \sum_{j=1}^m a^{\dagger}_{\text{out}, j} U_{ji} \,.
\end{equation}
Note that total photon number is conserved since linear optical transformations do not mix creation and annihilation operators. Applying this evolution rule to the Fock state $\kket{\bm{n}}$ yields (we henceforth work exclusively with output operators and drop the subscript):
\begin{equation}
	\label{eq:state_evolution}
    \frac{\left( \sum_{j_1=1}^m a_{j_1}^{\dagger} U_{j_1 1 } \right)^{n_1} 
    \ldots \left( \sum_{j_m=1}^m a_{j_m}^{\dagger} U_{j_m m} \right)^{n_m}}{\sqrt{n_1! \, n_2! \, ... \, n_m!}} \kket{\Omega} \,.
\end{equation}
This expression establishes that the state evolution may be computed by multiplying a sequence of homogeneous polynomials defined over the creation operators---each term in parentheses corresponds to a particular such term. Efficient strong simulation of linear optical circuits (meaning the simulation of the full statevector) may therefore be implemented by fast polynomial multiplication algorithms. Alternatively, the state evolution may also be described in terms of matrix permanents \cite{aaronson_computational_2011}.

A simple counting reveals that the set of states reachable through such transformations is exponentially suppressed. For a system with $m$ modes, the Hilbert space dimension is $\binom{m + n - 1}{n}$. However, a general linear optical circuit corresponds to an $m \times m$ unitary transfer matrix with $m^2$ free real parameters. As a result, the dimension of the Fock space is typically much greater than the dimension of the transfer matrix. An important consequence is that a given $n$-photon target state is unlikely to be reachable by a linear circuit acting upon a single Fock state.

This motivates the approach of heralded state preparation, which uses measured ancilla modes to conditionally prepare a target quantum state in the remaining unmeasured modes. When specific measurement outcomes are observed, they ``herald'' that the signal modes contain the desired entangled state. We assume that the modes are divided into two groups: $m_S$ \textit{signal} modes, and $m_A$ \textit{ancilla} modes. The ancilla modes will be measured at the end of the circuit, and the result will be used to post-select onto desired states. That is, a state $\kket{\Psi}$ over the full Fock space (signal and ancilla modes) may be decomposed in terms of ancilla Fock states:
\begin{equation}
    \label{eq:conditional_decomposition}
    \kket{\Psi} = \sum_{\bm{n}_A} \sqrt{\text{Pr}(\bm{n}_A)} \kket{\psi_{\bm{n}_A}}_s \otimes \kket{\bm{n}_A}_a \,.
\end{equation}
Here, $\kket{\bm{n}_A}_a$ is the Fock state over the ancilla photon counts $\bm{n}_A = (n_1, ..., n_{m_A})$, and $\kket{\psi_{\bm{n}_A}}_s$ is the corresponding state over the signal modes. The sum runs over all ancilla measurement tuples relevant for $m_A$ ancilla modes and $n$ total photons. We assume that this state is normalized by introducing the compensatory factor $\sqrt{\text{Pr}(\bm{n}_A)}$. The significance of this decomposition is that when the ancilla modes are measured and the outcome is $\bm{n}_A$, the signal modes are known to be in the state $\kket{\psi_{\bm{n}_A}}_s$, and this outcome occurs with probability $\text{Pr}(\bm{n}_A)$. With these conventions established, we now present our automated discovery pipeline that leverages this framework to optimize photonic circuits for heralded state preparation.

\section{Photonic circuit optimization pipeline \label{sec:optimization}}

\begin{figure*}[!htbp]
\includegraphics[width=\textwidth]{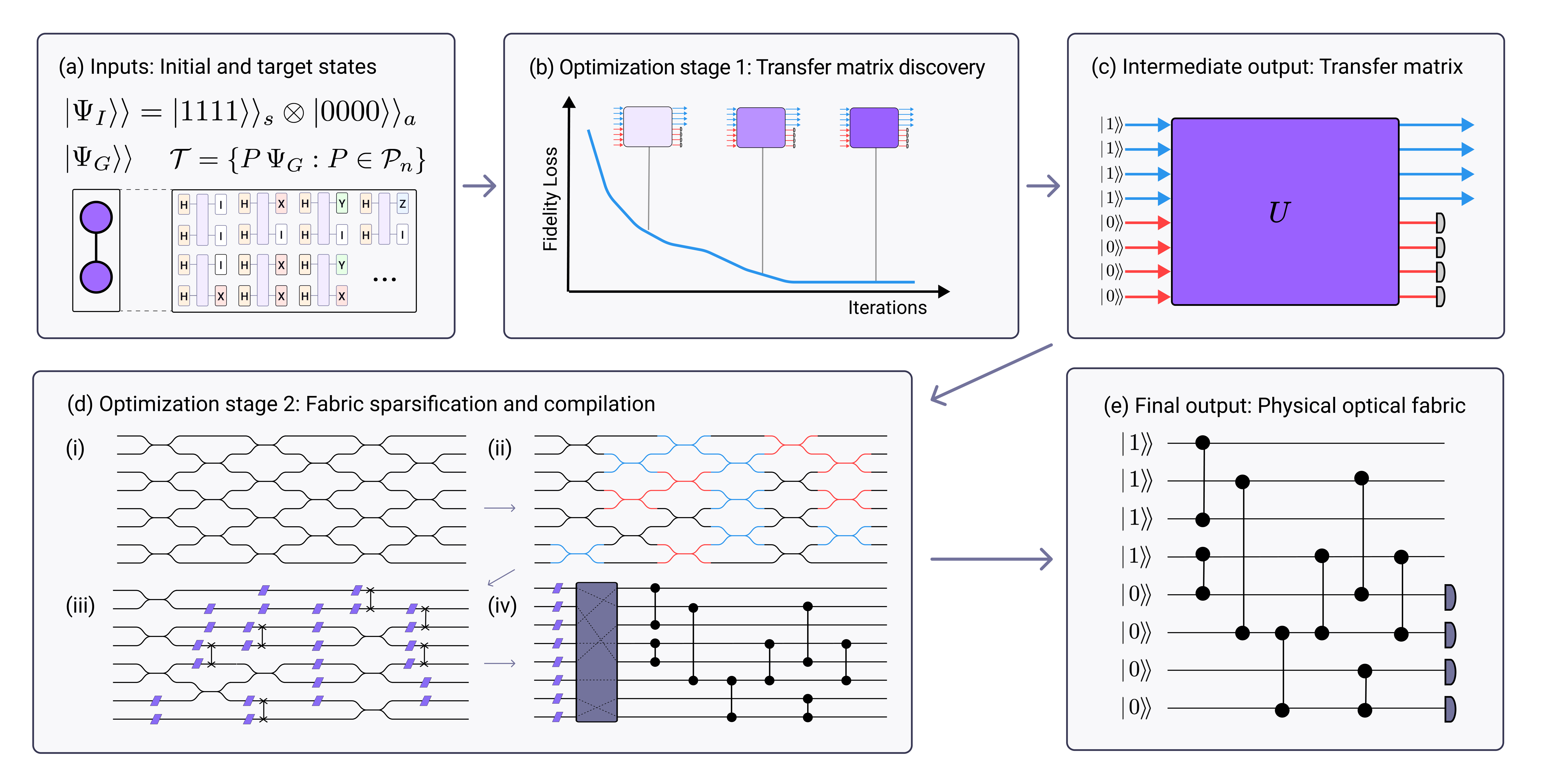}
    \caption{
    \label{fig:pipeline}
    Multi-stage optimization pipeline: 
(a) The pipeline receives as input one or more target states (in this case, a single 2-qubit graph state $\kket{\Psi_G}$) and an initial Fock state $\kket{\Psi_I}$, which utilizes additional modes and photons as ancillary resources. The single target state is optionally expanded into an equivalence class of target states $\mathcal{T}$---for instance those related to the original through the Pauli group or single-qubit Clifford operations. 
(b) Stage 1 optimizes a parameterized transfer matrix to maximize both the fidelity of the heralded output states with respect to the target state(s) and the overall success probability. 
(c) Intermediate output: the optimized unitary transfer matrix produced by the previous stage.
(d) Stage 2 seeks a simplified circuit that reproduces the same set of heralded output states.  
(i) Initial dense beamsplitter fabric obtained by applying the Clements {\em et al.} decomposition to the transfer matrix returned from Stage 1.  
(ii) A second optimization pass uses a regularized loss function to produce a sparsified fabric. Trivial beamsplitters (purely diagonal $2 \times 2$ transfer matrices) are shown in blue, and SWAP-equivalent beamsplitters (purely off-diagonal $2 \times 2$ transfer matrices) are shown in red.  
(iii) Decomposition of trivial and SWAP-equivalent beamsplitters into phaseshifters (purple quadrilaterals) and mode-SWAPs (vertical lines capped by ``x'').  
(iv) Commutation identities move all phaseshifters and SWAPs to the front of the circuit, where SWAPs combine into an overall mode permutation (gray box).  
(e) The final output is an experimentally realizable non-local optical fabric; beamsplitters may now act on non-adjacent modes, indicated by vertical lines capped by filled circles.
    }
\end{figure*}

Our approach to photonic circuit discovery combines fast, differentiable simulation with gradient-based optimization in a multi-stage pipeline, as depicted in Fig.~\ref{fig:pipeline}. The process begins by specifying the target quantum state and the available ancillary resources (number of additional modes and photons). Optionally, the single target state can be expanded to include all states related by single-qubit Pauli or Clifford operations. Both optimization stages rely on a common subroutine: differentiable strong simulation via polynomial multiplication. In the first stage, we optimize a unitary transfer matrix to produce the desired target state(s) with perfect fidelity and maximal success probability. In the second stage, we sparsify a circuit that implements the optimized transfer matrix by minimizing a regularized loss function that penalizes circuit complexity while preserving state fidelity and success probability. Finally, the sparsified circuit is compiled into an interpretable, hardware-friendly form suitable for experimental implementation. The following sections describe each component of this pipeline in detail.

\subsection{FFT-based strong simulation \label{sec:simulation}}
The task of numerically optimizing a photonic circuit to prepare a target state requires \textit{strong simulation} \cite{aaronson_computational_2011}, the computation of the full quantum state, as a subroutine. This is in contrast to weak simulation, also known as sampling, wherein the state is sampled in the Fock basis. Strong simulation is equivalent to the calculation of the permanent of a complex matrix, which is $\# P$-hard \cite{aaronson_computational_2011, valiant_complexity_1979}. Therefore, one approach towards strong simulation for linear optics quantum computing (LOQC) is to utilize existing algorithms for the computation of permanents. However, these approaches are unable to capitalize on certain simplifications that occur for the complex matrices that arise in the context of LOQC (namely, the existence of repeated rows in the matrix). Recently, a framework for strong simulation of LOQC systems was proposed in \cite{heurtel_strong_2023}. This framework, dubbed SLOS (for Strong Linear Optical Simulator) does not rely on permanents, and improves upon the previous state-of-the-art time complexity, at the expense of an increased memory requirement. This framework has been made available via the open-source \textit{Perceval} package \cite{heurtel_perceval_2023}. (More recently, Ref.~\cite{kolarovszki_piquasso_2025} introduced \textit{Piquasso}, another simulation tool that supports automatic differentiation.)

For our purposes, we require both the ability to perform strong simulation of a linear optical quantum circuit and to compute gradients of scalar-valued loss functions that quantify deviations from a target state. These requirements motivate a key computational innovation: a strong simulation algorithm compatible with automatic differentiation, enabling efficient gradient computation via backpropagation. To accomplish this, we developed a polynomial-based strong simulation algorithm and implemented it in \textit{PyTorch}, a popular machine learning framework \cite{paszke_pytorch_2019}.

A photonic state may be represented as a multivariate polynomial over the creation operators via the correspondence $\kket{\Psi} = P( \bm{a}^{\dagger} ) \kket{\Omega}$, where
\begin{equation}
    \label{eq:polynomial_representation}
    P(\bm{a}^{\dagger}) = \sum_{n_1=0}^{n} \ldots \sum_{n_m=0}^n T_{\bm{n}} \, (a_{1}^{\dagger})^{n_1} \ldots (a_{m}^{\dagger})^{n_m} \,,
\end{equation}
where $\bm{a}^{\dagger}$ refers to the set of creation operators over the $m$ modes. The polynomial coefficients $T_{\bm{n}}$ are related to the Fock state coefficients $c_{\bm{n}}$ appearing in Eq.~\ref{eq:general_state} via the normalization constant,
\begin{equation}
    \label{eq:coefficient_relation}
    T_{\bm{n}} = \frac{c_{\bm{n}}}{\sqrt{n_1! \ldots n_m !}} \,.
\end{equation}
Throughout this work, we assume an $n$-photon Fock state input. This, together with the photon conservation property of linear optical circuits, implies that the polynomial be homogeneous with degree $n$---in other words, $T_{\bm{n}} = 0$ for all photon number configurations that fail to satisfy $\sum_{i=1}^m n_i = n$ (or equivalently, for $\bm{n} \notin \Phi_{m,n}$). 

The polynomial corresponding to the output state is proportional to $\prod_{i=1}^m \left( \sum_{j=1}^m a_j^{\dagger} U_{ji} \right)^{n_i}$, a product over $m$ polynomials, with the degree of each term given by the number of photons in the corresponding input mode ($n_i$). This product may be computed using an FFT-based algorithm that exploits the correspondence between multivariate polynomial multiplication and multi-dimensional convolution. Specifically, the coefficients of a product of two multivariate polynomials can be computed as the convolution of their coefficient tensors. This observation forms the basis for our algorithm. We give a brief overview of the approach below; for further details, see Appendix~\ref{app:FFTdetails} (and also Ch. 8 of \cite{von_zur_gathen_modern_2013}).

The degree-$n$ output polynomial is numerically represented as a complex-valued tensor $T$ of shape $(n+1)^m$. The entries of this tensor are simply the polynomial coefficients, so that $T_{n_1 n_2 \ldots n_m} = T_{\bm{n}}$. To compute $T$, we introduce intermediate tensors $T^{(i)}$ that represent the contribution from each input mode. Specifically, $T^{(i)}$ corresponds to the polynomial representation of the transformed creation operator $\sum_{j=1}^{m} a_j^{\dagger} U_{ji}$ for input mode $i$. This tensor is given by
\begin{equation}
    \label{eq:intermediate_tensor}
    T^{(i)}_{n_1,\ldots,n_m} := \sum_{j=1}^{m} U_{ji} \, \mathbbm{1}_{\{n_j = 1 \text{ and } n_k = 0 \text{ for all } k \neq j\}} \,,
\end{equation}
where $\mathbbm{1}_{\{\cdot\}}$ is the indicator function (or Iverson bracket), which places matrix element $U_{ji}$ at the tensor position $(0,\ldots,0,1,0,\ldots,0)$ with the 1 in the $j$-th coordinate. The decomposition into these intermediate tensors enables the use of FFT-based convolution for efficient polynomial multiplication. The tensor for the output polynomial may be computed via
\begin{equation}
    \label{eq:PolyMulFFT}
    T = \frac{\mathrm{FFT}^{-1} \left[ \bigodot_{i=1}^{m} \mathrm{FFT} \left[ T^{(i)} \right]^{n_i} \right]}{\sqrt{n_1! \ldots n_m!}},
\end{equation}
where $\odot$ denotes pointwise tensor multiplication and the exponentiation denotes repeated pointwise multiplication of the tensor with itself. The normalization factor corresponds to the Fock state coefficient $T_{\bm{n}}$ for the input state, $\kket{\bm{n}}$. Once the output tensor is computed, Fock state probabilities, and conditional states may be straightforwardly obtained using Eq.~\ref{eq:coefficient_relation}.

The space complexity of our FFT-based simulation algorithm, \textsc{PolyMulFFT}, scales poorly due to the dense tensor representation that stores all $(n+1)^m$ entries despite most being zero due to photon number conservation:
\begin{equation}
    \mathrm{Space} = O\left( (n+1)^m \right).
\end{equation}
To determine the time complexity, note that an $m$-dimensional FFT on a tensor of shape $(n+1)^m$ has the same cost as a 1D FFT on a vector of equivalent size (note that a 1D FFT on a tensor of size $N$ has cost $O( N \log N)$), yielding $\mathrm{Time}(\textsc{FFT}) = O\left( m (n+1)^m \log n \right)$. Since we perform $m$ forward FFTs and one inverse FFT,  the total runtime is
\begin{equation}
    \mathrm{Time} = O\left( m^2 (n+1)^m \log n\right).
\end{equation}

This scaling becomes prohibitive for the preparation of heralded dual-rail qubit states for $\gtrsim 4$-qubit systems. However, this limitation can be addressed by exploiting the structure of our target states. We are specifically interested in preparing states where all Fock basis components contain the same number of photons in the signal modes---for example, the 2-photon dual-rail Bell state $(\kket{1010}_s - \kket{0101}_s)/\sqrt{2}$. For such outcomes, the total number of photons in the signal modes is a fixed quantity, denoted $n_S^T$. Due to photon number conservation, the corresponding ancilla states must also have a fixed photon number, $n_A^T := n - n_S^T$. In general, however, the full output state $\kket{\Psi}$ will involve superpositions of Fock states with different photon number splits between signal and ancilla modes. In particular, the Fock space admits a decomposition as a union of disjoint product sets, each with fixed photon numbers on the signal and ancilla rails:
\begin{equation}
    \Phi_{m, n} = \bigcup_{n_S = 0}^n \Phi_{m_S, n_S} \times \Phi_{m_A, n - n_S} \,.
\end{equation}
Consequently, a general state consisting of a superposition of many Fock states admits the decomposition
\begin{equation}
	\label{eq:polynomial_sector_decomposition}
	P( \bm{a}^{\dagger} ) =  \sum_{n_S=0}^{n} P^{(n_S,\,n-n_S)}(\bm{a}^\dagger) \,,
\end{equation}
where each component $P^{(n_S,\,n-n_S)}$ is homogeneous of degree $n_S$ in the signal output creation operators and degree $n-n_S$ in the ancilla output creation operators. Therefore, the efficiency of the algorithm can be improved by considering a restricted version of strong statevector simulation in which only the subspace ${\Phi_{m_S, n_S^T} \times \Phi_{m_A, n_A^T}}$ is simulated, corresponding to the $n_S = n_S^T$ term in Eq.~\ref{eq:polynomial_sector_decomposition}. The resulting algorithm, which we dub {\textsc{BipartitePolyMulFFT}} is described in detail in Appendix~\ref{app:FFTdetails}, including an asymptotic scaling analysis. In the symmetric case where $m_S = m_A = m/2$ and $n_S^T = n_A^T = n/2$, the space and time complexities are:
\begin{equation}
	\mathrm{Space} = O\left(\left(\frac{n}{2}+1\right)^{m/2}\right) \,,
\end{equation}
\begin{equation}
	\mathrm{Time} = O\left(\frac{m^2 \, 2^n}{\sqrt{n}} \left(\frac{n}{2}+1\right)^{m/2} \log n\right) \,.
\end{equation}
The restricted simulation in \textsc{BipartitePolyMulFFT} naturally requires less memory than the full version. The change in time complexity is a bit more subtle: while the exponent of the leading term is halved ($m \rightarrow m/2$), an exponential prefactor $2^n$ is introduced, arising from the $\binom{n}{n_S}$ ways of partitioning input photons between signal and ancilla modes. Whether this trade-off is favorable depends on the asymptotic scaling regime. If $m$ is held fixed as $n$ grows, \textsc{BipartitePolyMulFFT} becomes slower than \textsc{PolyMulFFT} for large photon counts. Conversely, if $m$ scales proportionally with $n$, the $(n+1)^{m/2}$ term dominates and the reduced exponent makes \textsc{BipartitePolyMulFFT} asymptotically superior. For the 4- and 5-qubit systems studied here, the bipartite restriction provides substantial practical speedup---the unrestricted algorithm proved too memory-intensive and slow to be viable in practice.

Finally, we note that this algorithm is fully compatible with automatic differentiation. We implemented the complete framework in \textit{PyTorch} \cite{paszke_pytorch_2019}, which provides native GPU acceleration and automatic differentiation support, enabling efficient gradient computation for the loss functions used in our optimization procedures. Having established an efficient simulation method, we now turn to the optimization procedure that explores the design space of possible transfer matrices.

\subsection{Transfer matrix optimization via Lie algebra parameterization \label{sec:optimizationpass1}}
As noted earlier, the optimization is carried out in two stages: first, the unitary is optimized using a Lie algebra parameterization; second, the resulting solution is sparsified and compiled into a simplified circuit.

The transfer matrix is parameterized using the Lie generators of the special unitary group SU$(m)$. Denoting these as $T_a$, for $a=1,..., m^2-1$, a general unitary may be written as 
\begin{equation}
    \label{eq:lie_parameterization}
    U = \exp\left(i \sum_a \xi_a T_a \right) \,,
\end{equation}
where $\xi_a$ are real coefficient parameters. This leads to a continuous optimization problem over the $\xi_a$. This approach is in contrast to alternative methods for circuit discovery that rely on discrete circuit topologies, and which necessarily suffer from the combinatorial explosion of all possible circuits for a given target state and initial state. Instead, here we search over a complete parametrization of the transfer matrix.

Next, we describe the loss function used to guide the automated discovery procedure. Recalling the conditional decomposition of the full signal + ancilla state (Eq.~\ref{eq:conditional_decomposition}), the loss should encourage the optimizer to find transfer matrices that (i) yield the desired target state(s) over the signal modes with near-perfect fidelity, and (ii) maximize the probability of generating those states.

The key components of the loss are: a distance measure over the space of signal states (rewarding proximity to the target), an aggregation over multiple target states, and an aggregation over the conditional states indexed by $\bm{n}_A$:\begin{equation}
    \label{eq:loss_function}
    \ell(U) = - \sum_{\bm{n}_A} \text{Pr}(\bm{n}_A) \min_{\psi_T \in \mathcal{T}} \vartheta_B(\psi_{\bm{n}_A}, \psi_T) \,,
\end{equation}
where $\mathcal{T}$ is the set of target states and $\vartheta_B$ denotes the Bures angle,
\begin{equation}
    \label{eq:bures}
    \vartheta_B(\psi_1, \psi_2) := \cos^{-1} \left( \frac{\left| \bbrakket{\psi_1}{\psi_2} \right|}{\sqrt{\bbrakket{\psi_1}{\psi_1} \bbrakket{\psi_2}{\psi_2}}} \right) \,.
\end{equation}
The Bures angle takes values in $[0, \pi/2]$, with $\vartheta_B = 0$ for physically-equivalent states, and $\vartheta_B = \pi/2$ for orthogonal states. In this formulation, the closest target state as measured by the Bures angle is selected for each conditional signal state via the $\min$ operation. The total loss is then computed as a probability-weighted sum over all conditional outcomes. To address vanishing gradients and improve convergence, we considered variants of the above loss where the Bures angle is transformed as ${\vartheta_B \rightarrow c \, (1-\vartheta_B)^{p}}$, where $c$ is an overall scale factor, and $p$ is a positive odd integer. A similar transform was used in \cite{stanisic_generating_2017}. 

This completes the first stage of our optimization pipeline. Successful optimization of the Lie algebra parameterization with Bures-angle-based loss produces a dense transfer matrix $U_1$ that prepares the target state(s) with high fidelity when specific ancilla patterns are observed. While implementing $U_1$ as a beamsplitter fabric would require up to $m(m-1)/2$ beamsplitters---acceptable for fixed parametric fabrics---discrete-component architectures significantly benefit from reduced beamsplitter count due to cumulative losses and fabrication complexity. The next stage addresses this by discovering a sparse circuit that preserves the state preparation capability with fewer optical elements.

\subsection{Circuit sparsification and compilation \label{sec:optimizationpass2}}

The output of the first optimization stage described above is an $m$-mode unitary transfer matrix $U_{1}$, which, when applied to an initial Fock state, produces an output state $\kket{\Psi_{1}}$. The goal of the second optimization stage is to represent $U_{1}$ as a photonic circuit that is as simple as possible, making it more amenable to experimental realization. While we do not model specific sources of experimental noise or decoherence, we incorporate practical considerations by minimizing the number of nontrivial optical components---an approach that generally reduces optical depth and may improve robustness.

In particular, we focus on minimizing the number of nontrivial, non-SWAP-equivalent beamsplitters in the circuit. A beamsplitter is considered trivial if its $2 \times 2$ transfer matrix is diagonal, and SWAP-equivalent if it is purely off-diagonal. These operations are excluded from the circuit complexity count, as trivial beamsplitters can be represented as two phaseshifters, and SWAP-equivalent beamsplitters decompose into two phaseshifters and a mode SWAP operation that can be absorbed via an overall permutation of the modes.

We accomplish the simplification of the unitary through a second optimization pass applied to $U_1$. First, the decomposition from Clements {\em et al.} \cite{clements_optimal_2016} is used to represent the previously optimized unitary as a beamsplitter fabric, with the fabric parameters collectively denoted as $f$: 
\begin{equation}
    \label{eq:beamsplitter_params}
    f = \{\theta^{(j)}, \phi_R^{(j)}, \phi_T^{(j)}\}_j \cup \{\varphi^{(i)} \}_{i=1}^m \,.
\end{equation}
Each beamsplitter is parameterized in terms of three angles, $(\theta, \phi_R, \phi_T)$. The first term  in Eq.~\ref{eq:beamsplitter_params} represents the union over all beamsplitter angles, while the second term represents the set of $m$ phaseshifter angles $\varphi$ (one for each mode). See Appendix~\ref{app:opticalconventions} for further details, including our conventions for beamsplitters and phaseshifters.

\begin{figure}[!htbp]
\centering
\includegraphics[width=1.0\linewidth]{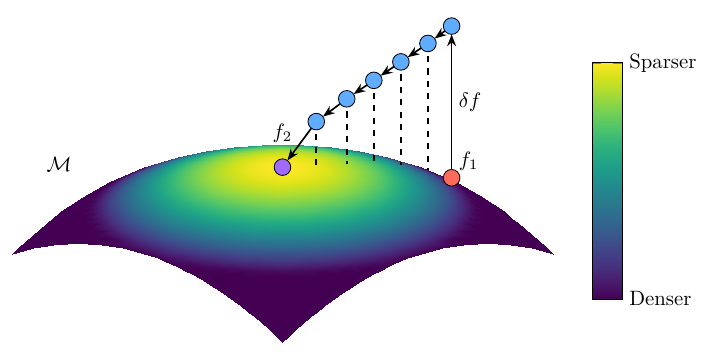}
\caption{
Notional illustration of fabric sparsification through regularized optimization. The optimization process simultaneously minimizes two objectives: the number of non-trivial beamsplitters and the deviation from perfect fidelity. The constraint surface $\mathcal{M}$ represents all fabric configurations that achieve zero fidelity loss. We initialize with fabric parameters $f_1$ obtained from the Clements {\em et al.} decomposition of the transfer matrix from the first optimization pass. By definition, $f_1$ lies on the constraint surface. To facilitate optimization, a small random perturbation $\delta f$ is applied, lifting the starting point away from the constraint surface. The optimization then proceeds through gradient-based iterations, with each step reducing the regularized loss function. The color scale indicates the sparsity level, with warmer colors corresponding to fewer beamsplitters. Dashed vertical lines illustrate the distance of each iteration from the constraint manifold $\mathcal{M}$, while solid arrows trace the optimization trajectory from the initial configuration $f_1$ through intermediate steps to the final sparse solution $f_2$.
}
\label{fig:manifold}
\end{figure}

The fabric parameters of the optimized unitary $U_1$ are denoted $f_1$, whereas $f$ refers to a general fabric. The optimized parameters are relaxed and allowed to deviate, i.e., $f = f_1 + \delta f$, with $\delta f$ a small random perturbation. New fabric parameters are then determined by minimizing a regularized loss that penalizes the number of non-trivial and non-SWAP-equivalent beamsplitters, as well as any deviations from the final state:
\begin{equation}
    \label{eq:regularized_loss}
    \ell(f) := \ell_{\text{fidelity}}(f) + \lambda_{\text{reg}} \sum_{\theta \in f} \left| \sin\left( 2 \theta \right) \right| \,.
\end{equation}
The first term is a fidelity loss quantifying the deviation between the optimized final state $\Psi_1$ and the state corresponding to the parameterized fabric $\Psi(f)$. The second term is a regularization loss that penalizes beamsplitters with ${\theta \neq 0 \text{ mod } \pi/2}$---that is, beamsplitters whose transfer matrix is not purely diagonal or purely off-diagonal. The sum runs over the $\theta$ parameter in every beamsplitter in the fabric. The relative strength of the two terms is controlled by $\lambda_{\text{reg}}$, a regularization hyperparameter.

We now discuss the choice of fidelity loss function, $\ell_{\text{fidelity}}$. The regularization procedure should preserve the heralded target state outputs of the optimized transfer matrix, as well as their success probabilities. To accomplish this, we require that the projection of the state onto the subspace with exactly $n_S^T$ photons in the signal modes and $n_A^T = n - n_S^T$ photons in the ancilla modes remains unchanged. This subspace, denoted $\Phi_{m_S, n_S^T} \times \Phi_{m_A, n_A^T}$, was introduced in Sec.~\ref{sec:simulation}. 

Accordingly, the fidelity loss used during sparsification is defined as
\begin{equation}
    \ell_{\text{fidelity}}(f) = \vartheta_B\left(\Psi(f), \mathcal{P}_{\text{target}} \Psi_1 \right) \,,
\end{equation}
where $\mathcal{P}_{\text{target}}$ denotes the projection onto the target subspace, and $\vartheta_B$ is the Bures angle (c.f., Eq.~\ref{eq:bures}). This formulation allows the regularized state $\Psi(f)$ to deviate from $\Psi_1$ on states lying outside the subspace of interest---states that will not affect the success probability for generating the target state(s). The resulting optimization problem is schematically illustrated in Fig.~\ref{fig:manifold}. The constraint surface $\mathcal{M} = \{ f: \ell_{\text{fidelity}}(f) = 0\}$ corresponds to the set of all fabrics that produce the same set of heralded target states, each with the same success probabilities, as the optimized transfer matrix. Importantly, the transfer matrix itself is not fixed along this surface. Minimization of Eq.~\ref{eq:regularized_loss} encourages the fabric to lie on the constraint surface while also using as few non-trivial and non-SWAP equivalent beamsplitters as possible. The result of this regularization is a second optimized unitary, $U_2$, described by a fabric with parameters $f_2$, which in general will differ from the result of the first optimization pass, $U_1 \neq U_2$. 

Lastly, once the regularized beamsplitter fabric is obtained as the result of this second optimization pass, the fabric is compiled into a simplified circuit consisting of non-local operations using the procedure depicted in Fig.~\ref{fig:pipeline}~(d). This compilation procedure makes use of circuit identities to first decompose trivial and SWAP-equivalent beamsplitters in terms of phaseshifters and SWAPs, and to then commute these operations to the front of the circuit, where they are easily seen to act trivially on Fock input states. For more details, see Appendix~\ref{app:compilation}.

\subsection{Example: 5-mode Bell state generator \label{sec:5modeexample}}

\begin{figure}[!htbp]
\includegraphics[width=0.35\textwidth]{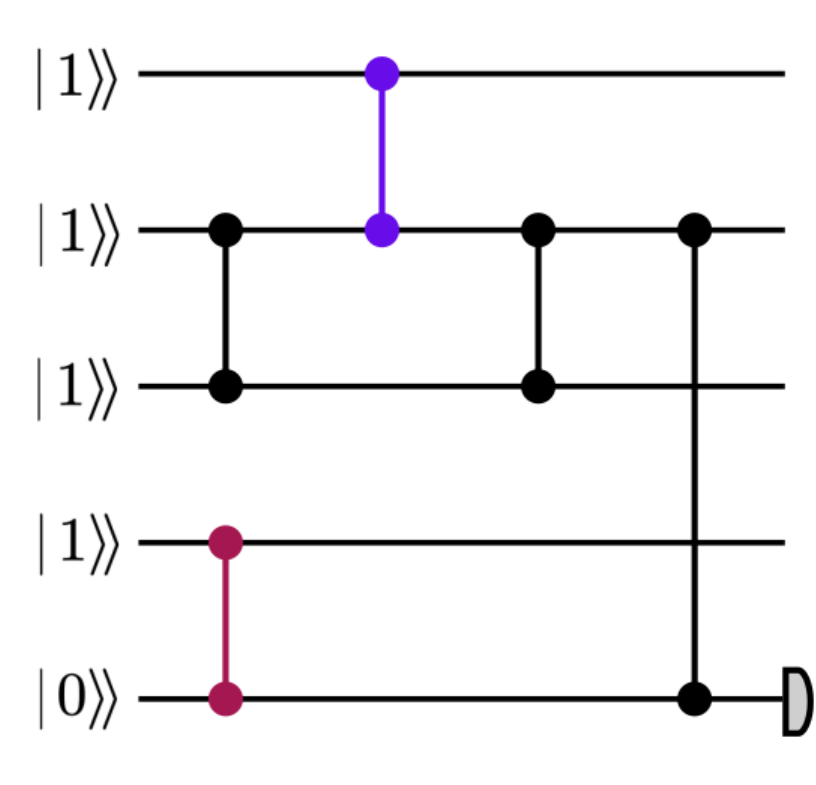}
    \caption{
    \label{fig:5mode_bsg}
    Automatically discovered 5-mode, 4 photon circuit for generating the dual-rail encoded Bell state $\kket{1010} - \kket{0101}$. Each horizontal line represents a photonic mode. The initial Fock state is shown to the left, with modes 1-4 receiving one input photon each. The circuit uses dual-rail encoding with mode pairs 1--2 and 3--4 each representing a logical qubit. CZ-gate symbols denote non-local beamsplitters that couple arbitrary mode pairs. Black beamsplitters are standard 50:50 couplers, the purple beamsplitter coupling modes 1-2 has reflection coefficient $R=2/3$, and the magenta beamsplitter coupling modes 4-5 has $R=1/3$. Note that the phaseshift angles $\phi_T$, $\phi_R$  are not specified here.
    } 
\end{figure}

As an illustrative example of our framework, we first consider the problem of generating a dual-rail Bell state using 5 modes and 4 photons. Fig.~\ref{fig:5mode_bsg} shows the result of our 2-stage optimization pipeline---a numerically-discovered and compiled circuit that produces a heralded Bell pair with success probability $1/9$. The modes are partitioned into four signal modes and a single ancilla mode, and the initial state is taken to be
\begin{equation}
    \kket{1111}_s \otimes \kket{0}_a = a_1^{\dagger} a_2^{\dagger} a_3^{\dagger} a_4^{\dagger} \kket{\Omega} \,.
\end{equation}
The circuit uses 5 beamsplitters: three with reflection coefficient $R=1/2$ (black, coupling modes (2,3) twice and (2,5) once)), one with $R = 2/3$ (purple, coupling modes (1,2)), and one with $R = 1/3$ (magenta, coupling modes (4,5)). The phase parameters $\phi_R$ and $\phi_T$ are not depicted for brevity. This circuit corresponds to the $5 \times 5$ transfer matrix
\begin{equation}
	\label{eq:BSGmatrix}
   U = \frac{1}{\sqrt{6}} 
   \begin{pmatrix}
    \sqrt{2} & \sqrt{2}\omega^6 & \sqrt{2} & 0 & 0 \\
    \omega^{18} & \omega^{16} & \omega^2 & \omega^6 & \sqrt{2} \\
    \sqrt{2}\omega^{18} & \sqrt{2}\omega^8 & \sqrt{2}\omega^{10} & 0 & 0 \\
    0 & 0 & 0 & 2 & \sqrt{2}\omega^6 \\
    \omega^6 & \omega^4 & \omega^{14} & \omega^6 & \sqrt{2}
    \end{pmatrix} \,,
    \end{equation}
where $\omega := e^{i\pi/12}$ is a 24-th root of unity, i.e. $\omega^{24} = 1$.

Applying the LOQC transformation rule, ${a_i^{\dagger} \rightarrow \sum_j a_j^{\dagger} U_{ji}}$, the input polynomial is mapped to the output polynomial
\begin{widetext}
\begin{align}
	\label{eq:BSGexample}
    a_1^{\dagger} a_2^{\dagger} a_3^{\dagger} a_4^{\dagger}  &\rightarrow
    -\frac{\sqrt{2} \, (a^{\dagger}_1)^{3} (a^{\dagger}_2)}{18}  
    + \frac{\sqrt{2} \, i (a^{\dagger}_{1})^{3} (a^{\dagger}_{4})}{9} 
    - \frac{\sqrt{2} \, (a^{\dagger}_{1})^{3} (a^{\dagger}_{5})}{18} 
    - \frac{(a^{\dagger}_{1}) (a^{\dagger}_{2})^{2} (a^{\dagger}_{3})}{6}
    + \frac{i (a^{\dagger}_{1}) (a^{\dagger}_{2}) (a^{\dagger}_{3}) (a^{\dagger}_{4})}{3} 
    + \frac{i (a^{\dagger}_{5})^{4}}{36} \\
    &- \frac{i (a^{\dagger}_{1}) (a^{\dagger}_{3}) (a^{\dagger}_{4}) (a^{\dagger}_{5})}{3} 
    + \frac{(a^{\dagger}_{1}) (a^{\dagger}_{3}) (a^{\dagger}_{5})^{2}}{6} 
    - \frac{i (a^{\dagger}_{2})^{4}}{36}
    - \frac{(a^{\dagger}_{2})^{3} (a^{\dagger}_{4})}{18}
    + \frac{i (a^{\dagger}_{2})^{3} (a^{\dagger}_{5})}{18}
    + \frac{(a^{\dagger}_{2})^{2} (a^{\dagger}_{4}) (a^{\dagger}_{5})}{6} \nonumber \\
    &- \frac{\sqrt{2} \, i (a^{\dagger}_{2}) (a^{\dagger}_{3})^{3}}{18}
    - \frac{(a^{\dagger}_{2}) (a^{\dagger}_{4}) (a^{\dagger}_{5})^{2}}{6} 
    - \frac{i (a^{\dagger}_{2}) (a^{\dagger}_{5})^{3}}{18} 
    - \frac{\sqrt{2} \, (a^{\dagger}_{3})^{3} (a^{\dagger}_{4})}{9} 
    - \frac{\sqrt{2} \, i (a^{\dagger}_{3})^{3} (a^{\dagger}_{5})}{18}
    + \frac{(a^{\dagger}_{4}) (a^{\dagger}_{5})^{3}}{18} \nonumber \,.
\end{align}
\end{widetext}

This expression corresponds to the polynomial representation of the output state, as in Eq.~\ref{eq:polynomial_representation}. The tensor representation of this output polynomial can be numerically computed using our FFT-based polynomial multiplication algorithm (Sec.~\ref{sec:simulation}). The tensors $T$, $T^{(i)}$ for $i=1, ..., m$ have shape $(5, 5, 5, 5, 5)$---a rank-$m=5$ tensor where each dimension has size $n+1=5$ to accommodate all possible photon number configurations from 0 to $n$ in each mode.

Each term in the output polynomial corresponds to a specific way the four input photons can be distributed across the five output modes after the linear optical transformation. To extract the desired Bell state, we implement the heralding protocol: we post-select on measurement outcomes that place exactly 2 photons in the ancilla mode (mode 5), which yields the term ${(a_{1}^{\dagger} a_{3}^{\dagger} - a_{2}^{\dagger} a_{4}^{\dagger}) (a_{5}^{\dagger})^{2} /6}$. Restoring the Fock state normalization factors (c.f. Eq.~\ref{eq:coefficient_relation}, in this case corresponding to an overall factor of $\sqrt{2}$), this results in the dual-rail state
\begin{equation}
    \frac{1}{3} \left( \frac{ \kket{1010}_s - \kket{0101}_s}{\sqrt{2}} \right) \otimes \kket{2}_a \,.
\end{equation}
The expression in parentheses is a properly normalized state on the signal modes, and thus the success probability is given by the squared amplitude prefactor: $|1/3|^2 = 1/9$ (cf. Eq.~\ref{eq:conditional_decomposition}). When the ancilla measurement yields exactly 2 photons, this heralds that the four signal modes contain the Bell state $(\ket{00} - \ket{11})/\sqrt{2}$ in dual-rail encoding.

A circuit functionally equivalent to that in Fig.~\ref{fig:5mode_bsg} was first reported in Fldzhyan {\em et al.} \cite{fldzhyan_compact_2021}. Our framework independently discovered this circuit, up to circuit rearrangements, mode relabellings, and choice of target Bell state (our solution was discovered using the Pauli group equivalence class, meaning that the target states $\mathcal{T}$ include all states related to the standard Bell state up to single-qubit Paulis). This independent recovery of a known circuit validates our approach's ability to navigate the exponentially large space of possible photonic circuits and identify high-quality solutions. The algorithm was provided only with the target state specification and circuit constraints (number of modes, photons, and initial state), demonstrating the method's capability for genuine circuit discovery. 

The 5-mode Bell state generator demonstrates the complete pipeline in action: from transfer matrix optimization through sparsification to final circuit compilation. The independent rediscovery of the Fldzhyan {\em et al.} circuit validates our approach, while the 1/9 success probability establishes a baseline for comparison with more complex graph states. Having illustrated the methodology with this concrete example, we now present comprehensive results for graph states up to 5 qubits, where the automated nature of our approach becomes essential for exploring the exponentially growing design space.

\section{Automated discovery of graph states \label{sec:graphstateresults}}

\begin{figure*}[!htbp]
\centering
\includegraphics[width=1.0\textwidth]{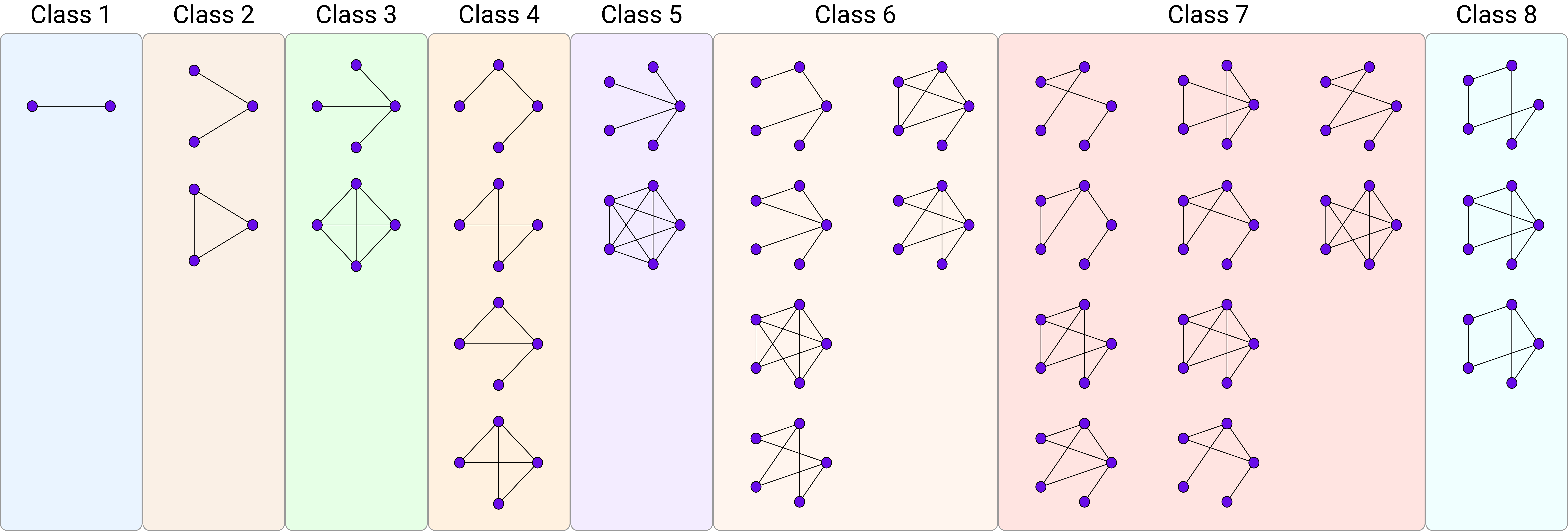}
\caption{
\label{fig:all_graphs}
All connected, non-isomorphic graphs with 2, 3, 4, and 5 nodes, organized by local-complementation (LC) equivalence classes. Graphs within each equivalence class can be transformed into one another via a sequence of local complementation operations, and the corresponding graph states can be transformed into each other via local single-qubit Clifford operations. The equivalence class numbering follows \cite{hein_multiparty_2004}.
}
\end{figure*}

Graph states serve as entangled resource states in measurement-based quantum computing (MBQC). A graph state is defined with respect to an undirected graph $G = (V, E)$, where each vertex represents a qubit initialized in state $\ket{+} = \frac{1}{\sqrt{2}}(\ket{0} + \ket{1})$, and each edge corresponds to a controlled-Z (CZ) gate applied between the connected qubits:
\begin{equation}
    |G \rangle = \prod_{(i,j) \in E} CZ_{i,j} \ket{+}^{\otimes |V|} \,.
\end{equation}
Graph states are stabilizer states with a unique stabilizer generator for each vertex $v \in V$: $K_v = X_v \prod_{u \in \mathcal{N}(v)} Z_u$, where $X_v$ is the Pauli-X operator on qubit $v$, $Z_u$ is the Pauli-Z operator on neighboring qubit $u$, and $\mathcal{N}(v)$ denotes the neighbors of vertex $v$ in graph $G$. The specialized term ``cluster state'' typically refers to graph states where the underlying graph forms a regular lattice. As the largest state we consider is 5 qubits, we use the more general terminology ``graph state'' throughout.

In this study, we examine graph states over 2, 3, 4, and 5 nodes, as illustrated in Fig.~\ref{fig:all_graphs}. Two key considerations shape our analysis. First, it suffices to consider connected graphs up to isomorphism, since graph states of disconnected graphs are tensor products of their connected components' graph states, and isomorphisms merely permute qubit labels. Second, in many quantum information contexts, stabilizer states that differ only by single-qubit Clifford operations are effectively equivalent. This motivates organizing graph states into equivalence classes under local Clifford transformations.

\begin{figure}[!htbp]
\includegraphics[width=0.5\textwidth]{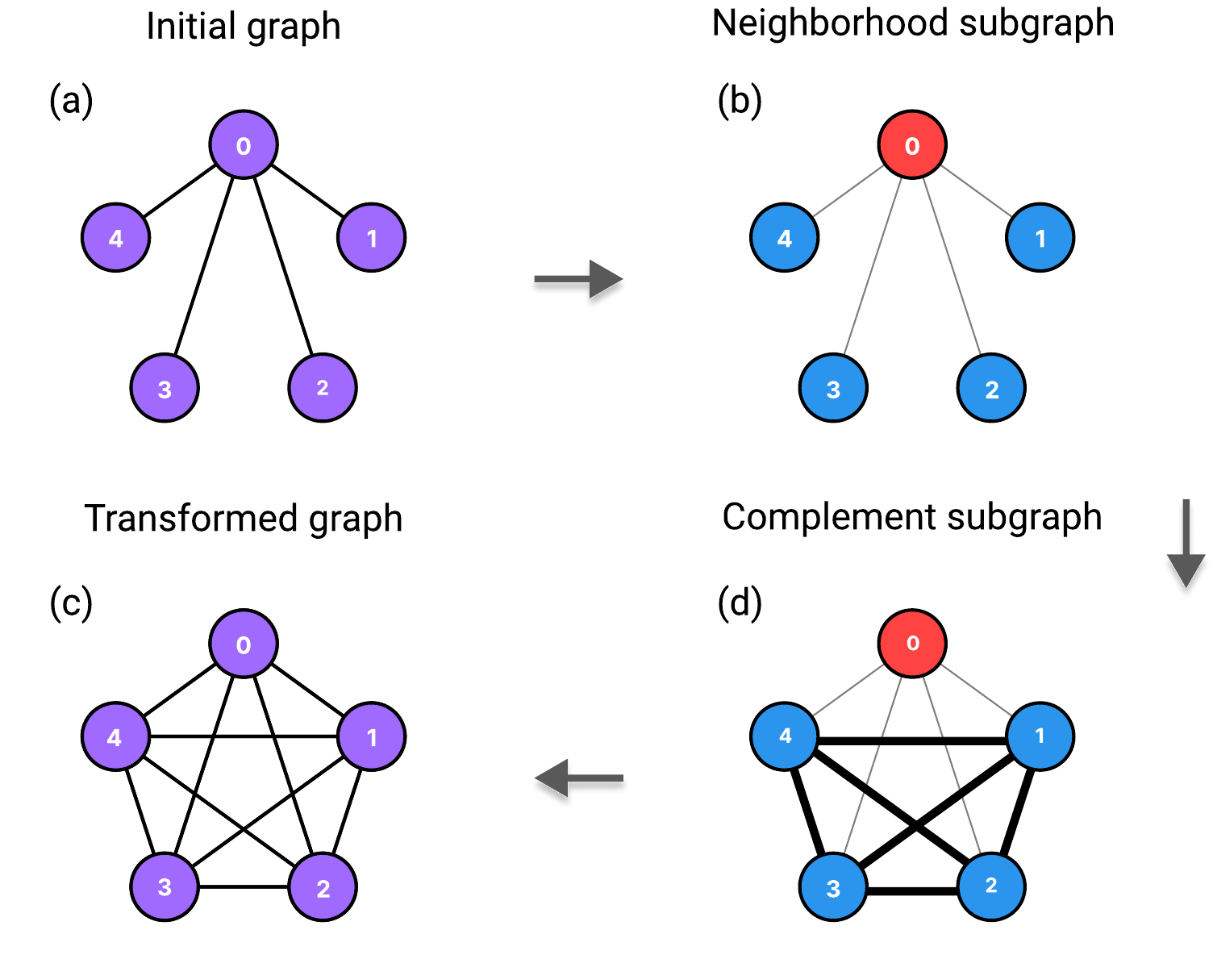}
    \caption{
    \label{fig:localcomplementation}
    Local complementation. (a) The initial graph is the star graph. (b) The local complementation operation is applied with respect to vertex 0 (shown in red). The neighborhood subgraph is the graph formed by restricting to the neighborhood of vertex 0, shown in blue. (c) The complement of the neighborhood subgraph is the fully connected graph among nodes 1-4. (d) The final, transformed graph replaces the neighborhood subgraph with its complement, resulting in a fully connected graph over 5 nodes.
    }
\end{figure}

Importantly, the relationship between graph states and their equivalence classes is not one-to-one: Although all stabilizer states can be represented as graph states up to local Clifford operations \cite{hein_multiparty_2004}, multiple graph states can belong to the same equivalence class. Two graph states are locally equivalent if their underlying graphs are related by a sequence of local complementation (LC) operations \cite{van_den_nest_graphical_2004, van_den_nest_efficient_2004}. Fig.~\ref{fig:localcomplementation} demonstrates this operation applied to a 5-node star graph, showing its LC-equivalence to the complete graph---meaning their corresponding graph states are locally equivalent. In Fig.~\ref{fig:all_graphs}, connected non-isomorphic graphs are organized by their LC equivalence classes, with all graphs in the same class representing locally equivalent states. For instance, the 2-node graph state (single edge) is locally equivalent to a Bell state (considered in Sec.~\ref{sec:5modeexample}), while star graph states are locally equivalent to both fully connected graph states and to GHZ states.

Next, we note that a circuit that generates a graph state $\ket{G}$ with unit fidelity under Pauli equivalence can be adapted to generate any LC-equivalent graph state $\ket{G'}$. Let $\mathfrak{M}$ denote the set of ancilla measurement outcomes for a given circuit, and let $\mathfrak{M}_{\mathcal{T}} \subseteq \mathfrak{M}$ denote outcomes that successfully herald the target graph state under Pauli equivalence. For each $\bm{n}_A \in \mathfrak{M}_{\mathcal{T}}$, define $P_{\bm{n}_A}$ as the Pauli operator that maps the heralded state into the target state, $P_{\bm{n}_A} \ket{\psi_{\bm{n}_A}} = \ket{G}$. Since $G \sim G'$ under local complementation, the corresponding graph states are related by a local Clifford: $\ket{G'} = C\ket{G}$, where $C$ is an $(m_S/2)$-qubit Clifford unitary, implementable as an $m_S$-mode linear optical transformation under dual-rail encoding. Being a local Clifford, $C$ decomposes into single-qubit Cliffords---each realizable via an $SU(2)$ beamsplitter on each dual-rail qubit. A naive implementation would apply $C$ after the measurement-dependent Pauli correction $P_{\bm{n}_A}$, but our model assumes only feedforward Pauli operations. We resolve this by exploiting Clifford conjugation: ${C P_{\bm{n}_A} = P_{\bm{n}_A}' C}$, allowing $C$ to be applied \textit{before} the conditional correction, at the cost of modifying the per-outcome Pauli operators: $P_{\bm{n}_A} \to P_{\bm{n}_A}'$ for each $\bm{n}_A \in \mathfrak{M}_{\mathcal{T}}$.

A second natural equivalence class to consider is equivalence under single-qubit Pauli rotations. The Pauli group forms a subgroup of the single-qubit Clifford group and is particularly relevant because Pauli corrections can often be tracked classically and applied as conditional feedforward operations based on measurement outcomes. Both equivalence classes can be incorporated into our numerical optimization by expanding the set of target states in the loss function (cf.~Eq.~\ref{eq:loss_function}). From a computational perspective, while both the Pauli and single-qubit Clifford groups grow exponentially in the number of qubits (as $4^n$ and $24^n$, respectively), we found that expanding the desired target state to the local Clifford equivalence class created significant computational bottlenecks for $n \geq 4$ qubits. As Table~\ref{tab:discovered_circuits} illustrates, the single-qubit Clifford equivalence class is 2-3 orders of magnitude larger than the Pauli equivalence class for the considered states. Therefore, in the results presented here, we focus exclusively on the Pauli equivalence class while noting that the framework naturally extends to include Clifford equivalences for smaller systems or when computational resources permit. Note that the size of the Pauli equivalence class is the same for all stabilizer states by virtue of the orbit-stabilizer theorem \cite{rotman_introduction_2012}. The Pauli group has $4^n$ elements, and the stabilizer group has order $2^n$, meaning that the orbit size (number of elements in the equivalence class) is $4^n / 2^n = 2^n$. 

\subsection{Results}

\newcommand{\graphwidth}{0.8cm}
\begin{table*}[!htbp]
\centering

\caption{Resource requirements for numerically discovered graph state preparation circuits. ``WL Hash'' serves as an identifier for each graph isomorphism equivalence class, and refers to the Weisfeiler-Lehman (WL) graph hash, as computed by the \texttt{networkx} package \cite{hagberg_exploring_2008}. (The WL hash is identical for isomorphic graphs, and typically non-isomorphic graphs will have different hash values, although this is not guaranteed \cite{shervashidze_weisfeiler-lehman_2011}.) Nodes are canonically labeled $V = \{0, 1, ..., n-1\}$, and the first 8 characters of the hash are reported. ``Success Prob'' refers to the success probability for the discovered circuit, the probability of observing one of the target states with fidelity $> 0.9999$. 
``States'' refers to the number of such heralded target states produced by the optimized circuit. ``BS'' refers to the final beamsplitter count of the sparsified and compiled circuit. ``Modes'' and ``Photons'' refer to the total number of optical modes and photons used in the circuit, respectively. ``Pauli Orbit'' refers to the size of the Pauli orbit (equivalence class) containing the target graph state. ``LC Class'' identifies the Local Complementation equivalence class using the convention of Ref.~\cite{hein_multiparty_2004}, and ``$\text{Cliff}_1^{\otimes n}$ Orbit'' counts the number of locally-equivalent stabilizer states under the action of the local Clifford group, $\text{Cliff}_1^{\otimes n}$.
}

\vspace{0.5em} %

\begin{tabular}{c|c|c|c|c|c|c|c|c|c|c|c}
\toprule
\textbf{Qubits} & \textbf{Graph} & \textbf{WL Hash} & \textbf{Success Prob} & \textbf{States} & \textbf{BS} & \textbf{Modes} & \textbf{Photons} & \makecell{\textbf{Pauli} \textbf{Orbit}} & \textbf{LC Class} & \makecell{\textbf{Cliff$_1^{\otimes n}$} \textbf{Orbit}} \\
\midrule
\multirow{3}{*}{3Q}
  & \includegraphics[valign=c,width=\graphwidth]{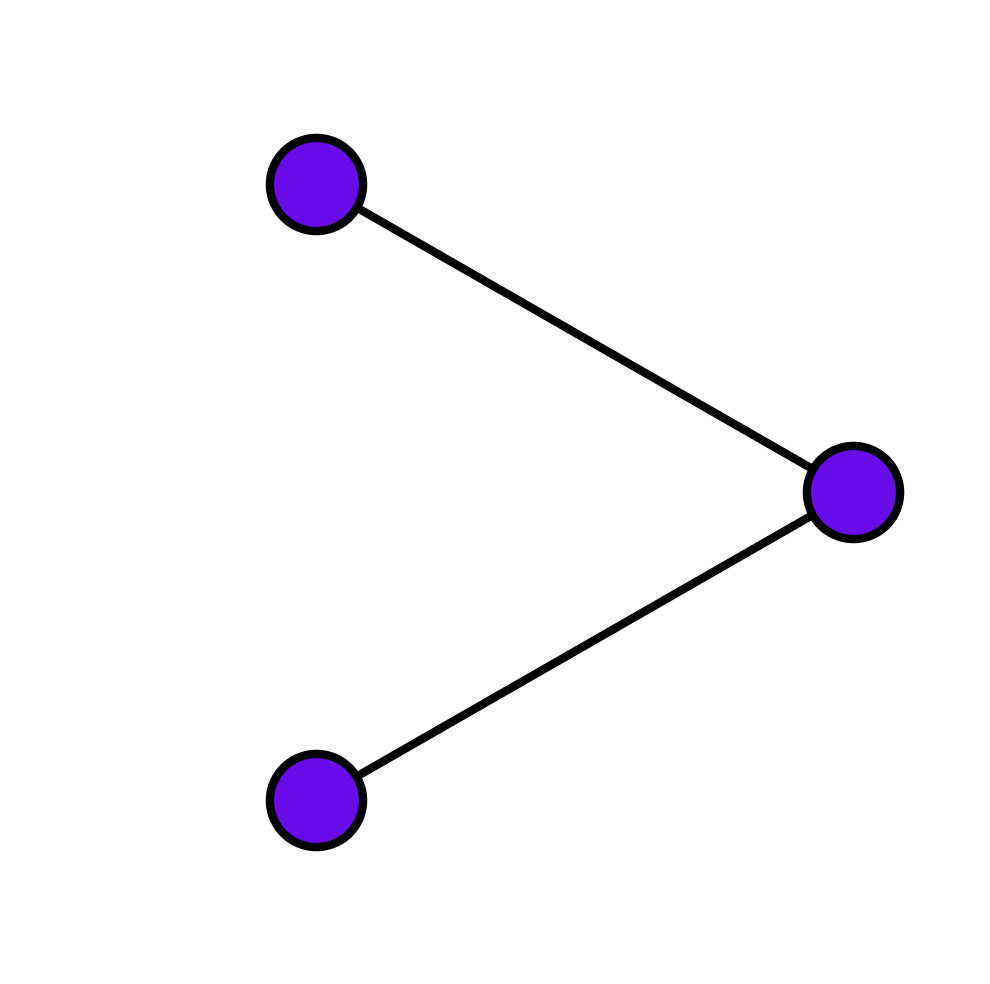} & \texttt{5144181a} & 1.852E-2 & 3 & 17 & 11 & 6 & 8 & 2 & 432 \\
  & \includegraphics[valign=c,width=\graphwidth]{figures/graph_3Q_5144181ac27497fdfa9bdb5b8b799630.png} & \texttt{5144181a} & 2.215E-2 & 12 & 29 & 15 & 10 & 8 & 2 & 432 \\
  & \includegraphics[valign=c,width=\graphwidth]{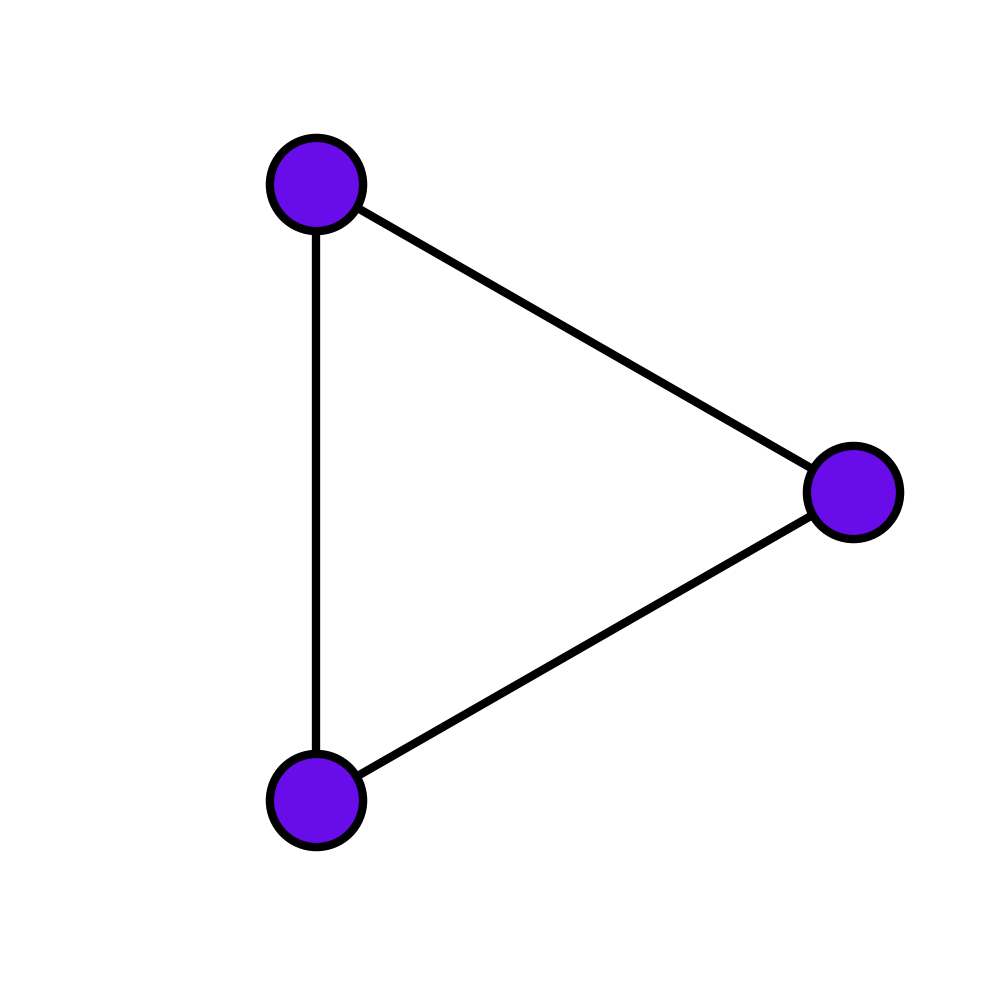} & \texttt{b4844a24} & 3.125E-2 & 8 & 23 & 12 & 6 & 8 & 2 & 432 \\
\midrule
\multirow{6}{*}{4Q}
  & \includegraphics[valign=c,width=\graphwidth]{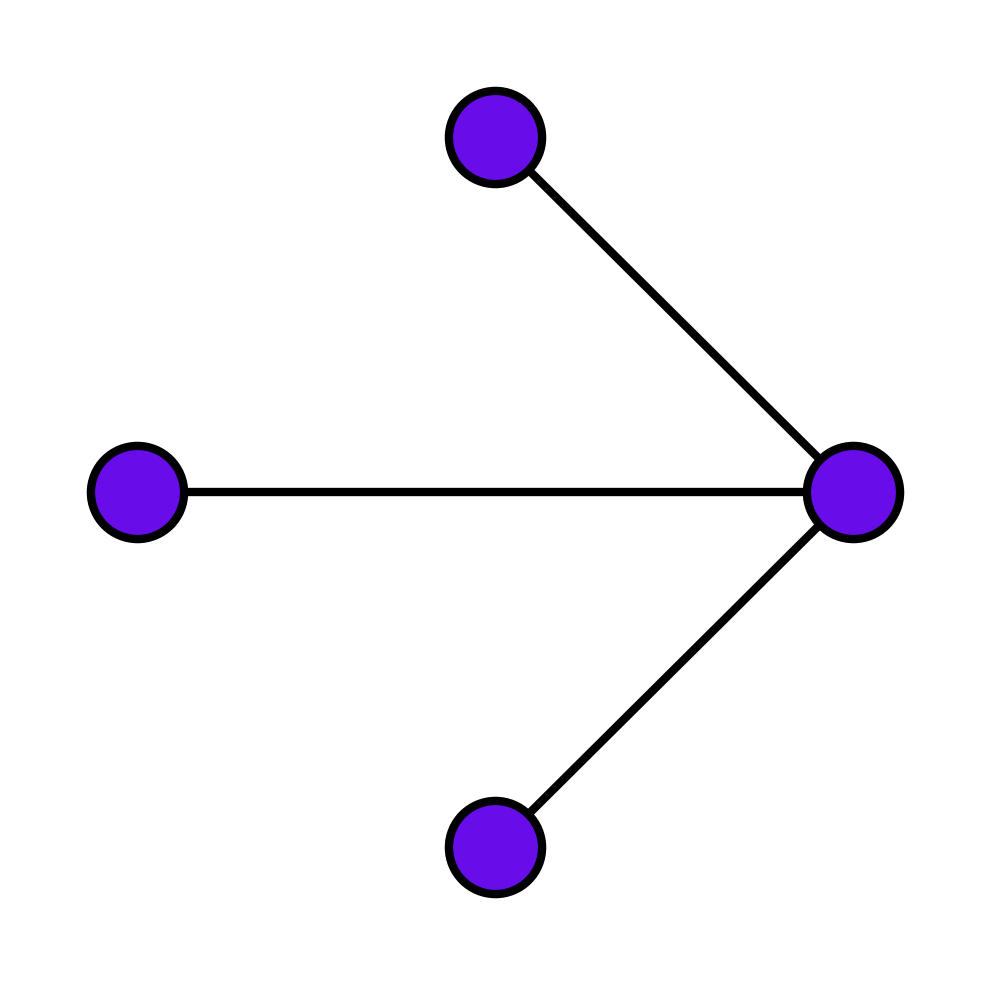} & \texttt{cde6b48e} & 4.630E-3 & 4 & 18 & 14 & 9 & 16 & 3 & 2592 \\
  & \includegraphics[valign=c,width=\graphwidth]{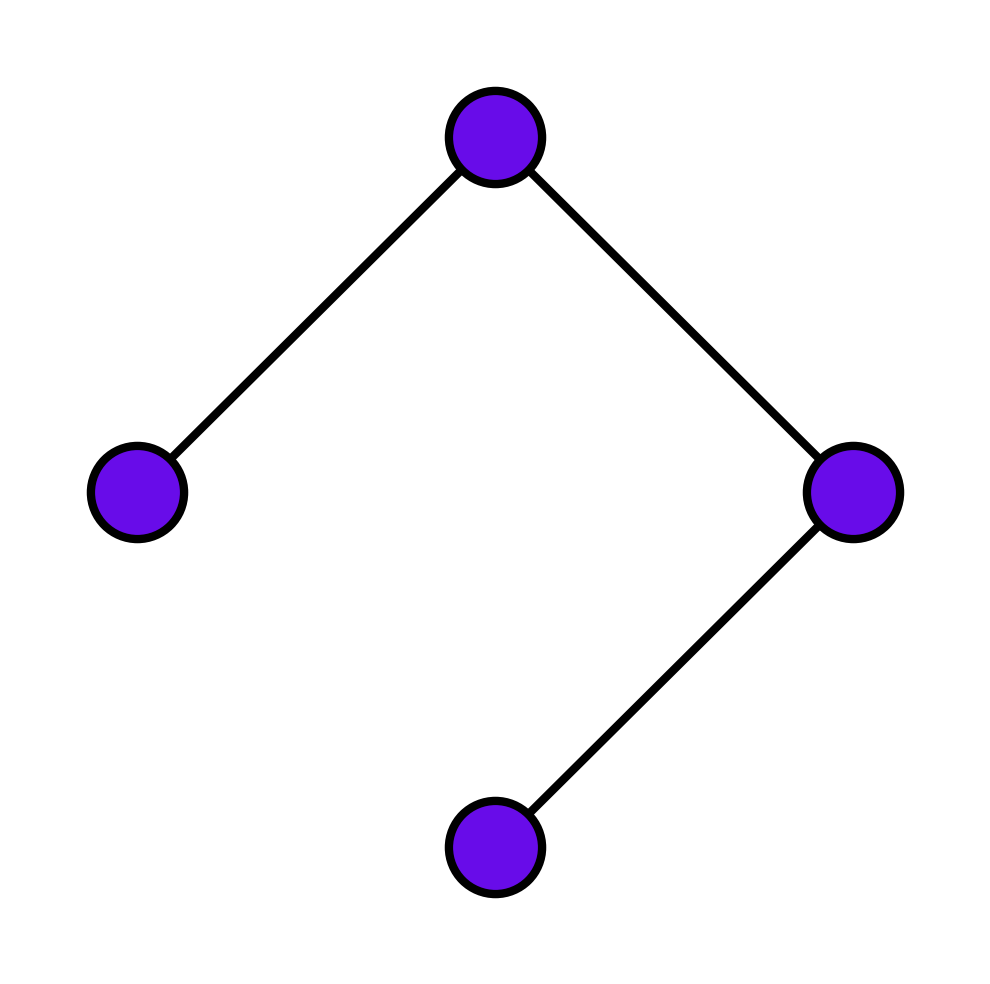} & \texttt{69144809} & 2.053E-3 & 2 & 45 & 15 & 9 & 16 & 4 & 5184 \\
  & \includegraphics[valign=c,width=\graphwidth]{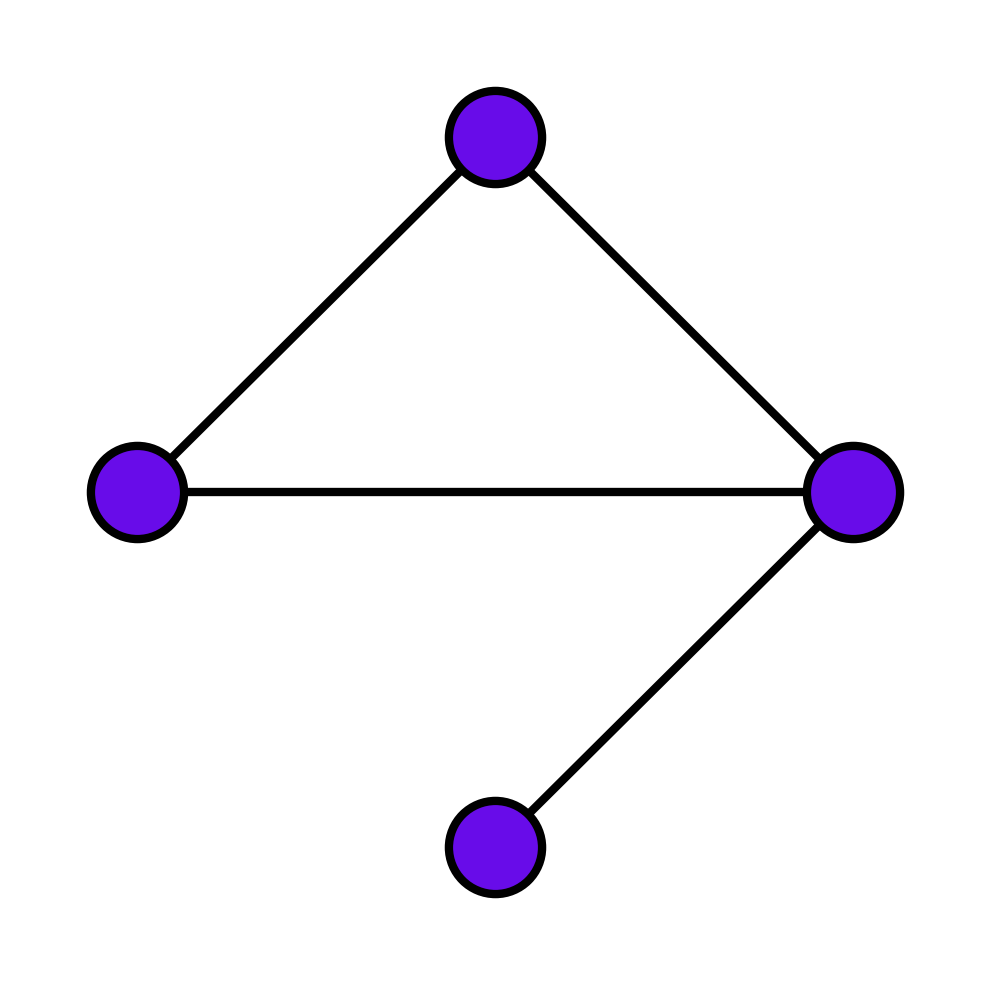} & \texttt{7bc4dde9} & 3.474E-3 & 8 & 26 & 15 & 9 & 16 & 4 & 5184 \\
  & \includegraphics[valign=c,width=\graphwidth]{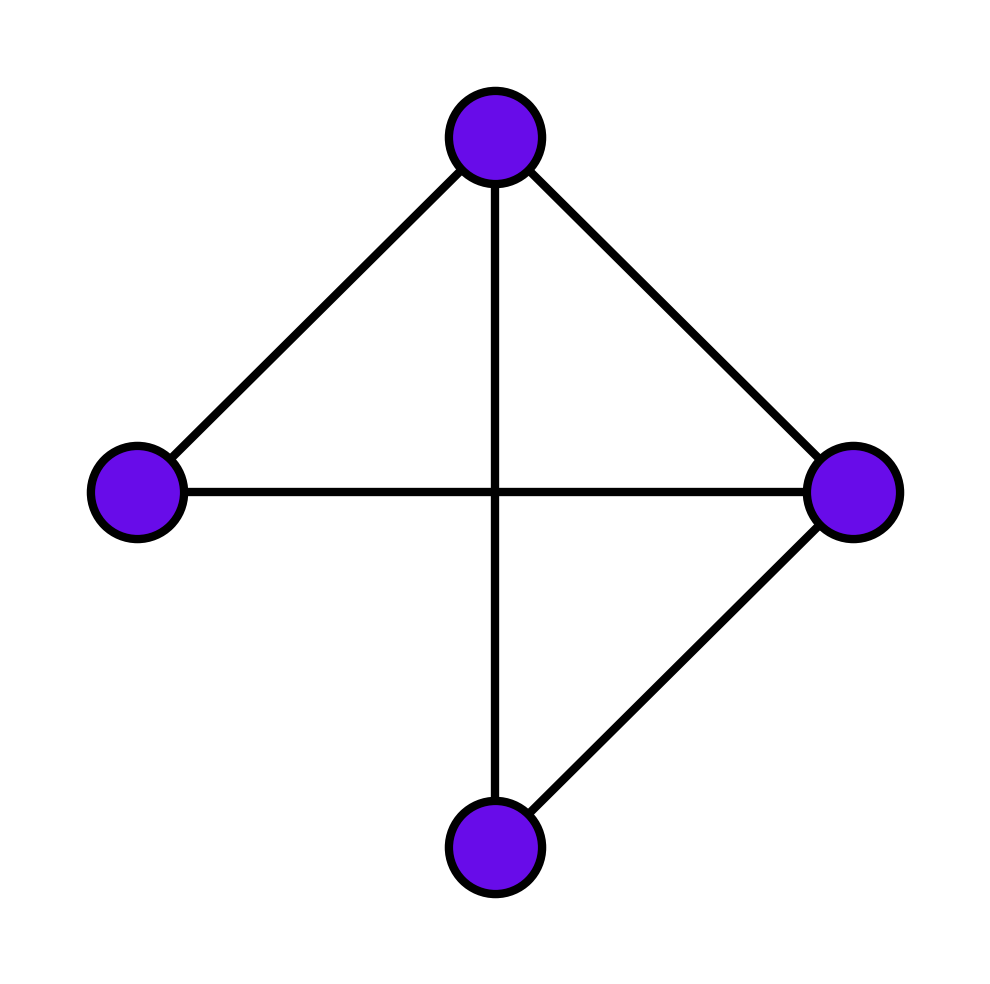} & \texttt{93141c35} & 3.579E-3 & 18 & 29 & 17 & 9 & 16 & 4 & 5184 \\
  & \includegraphics[valign=c,width=\graphwidth]{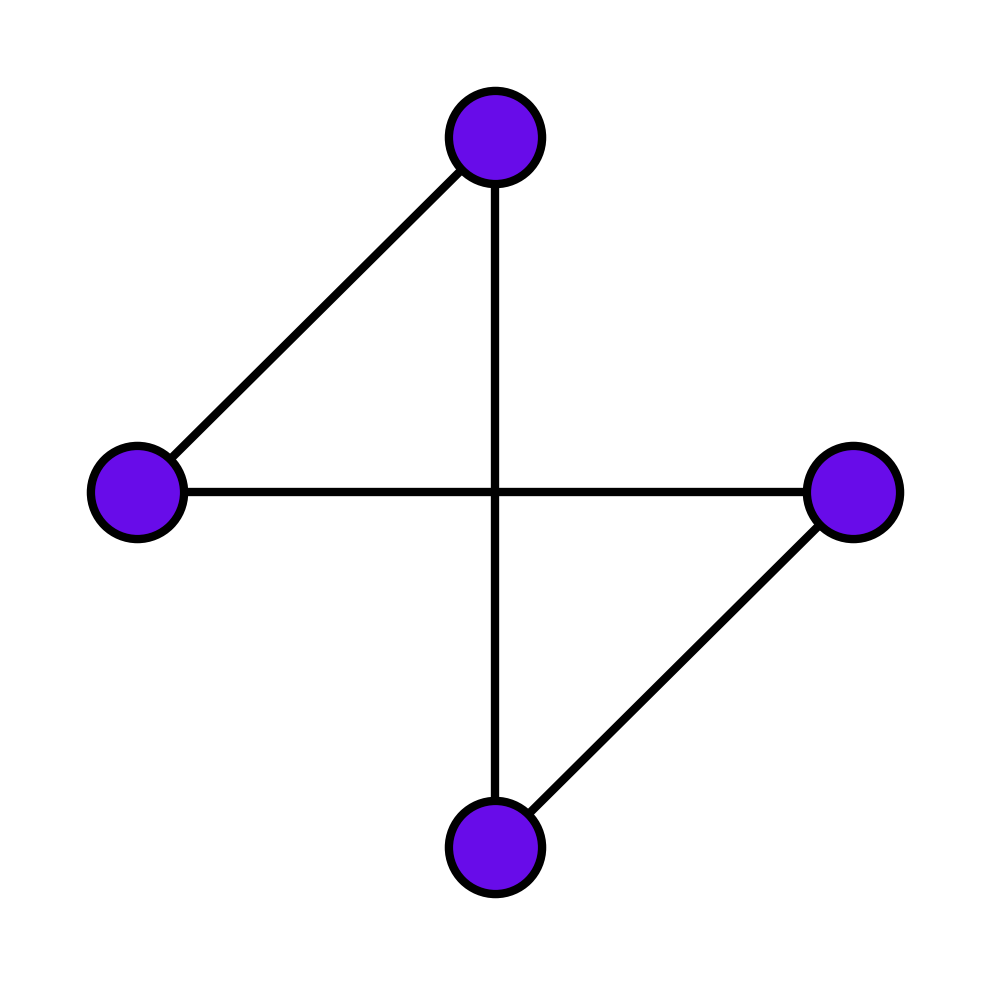} & \texttt{20a60ed0} & 2.066E-3 & 3 & 44 & 16 & 9 & 16 & 4 & 5184 \\
  & \includegraphics[valign=c,width=\graphwidth]{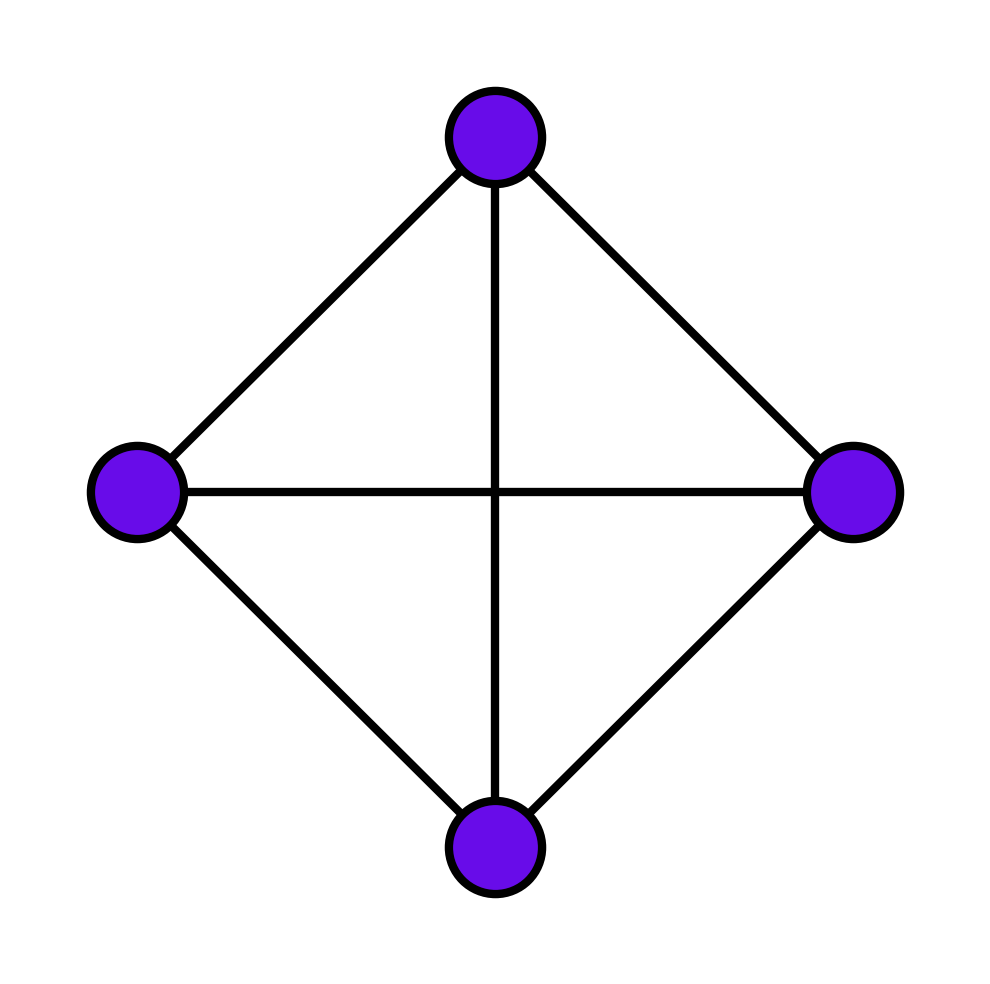} & \texttt{d2e3f71e} & 7.813E-3 & 24 & 23 & 17 & 8 & 16 & 3 & 2592 \\
\midrule
\multirow{3}{*}{5Q}
  & \includegraphics[valign=c,width=\graphwidth]{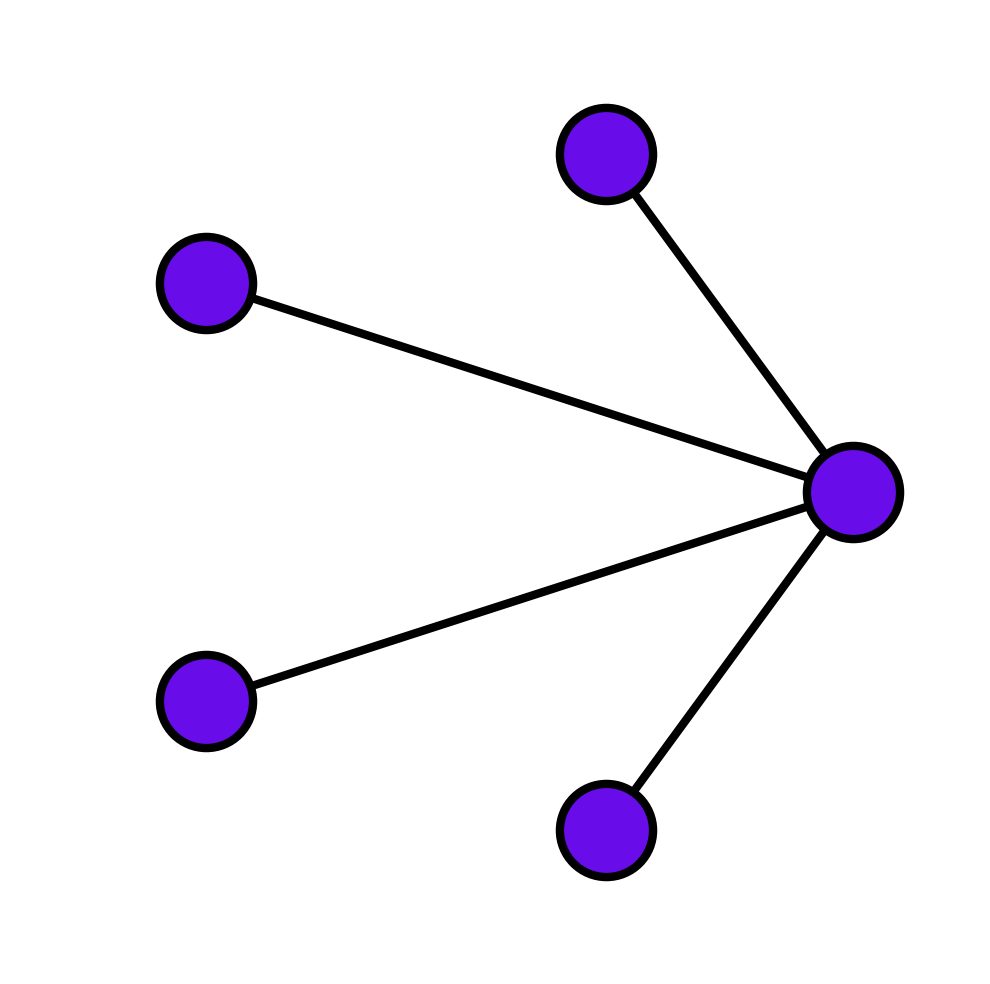} & \texttt{c502b67e} & 1.157E-3 & 12 & 33 & 19 & 10 & 32 & 5 & 15552 \\
  & \includegraphics[valign=c,width=\graphwidth]{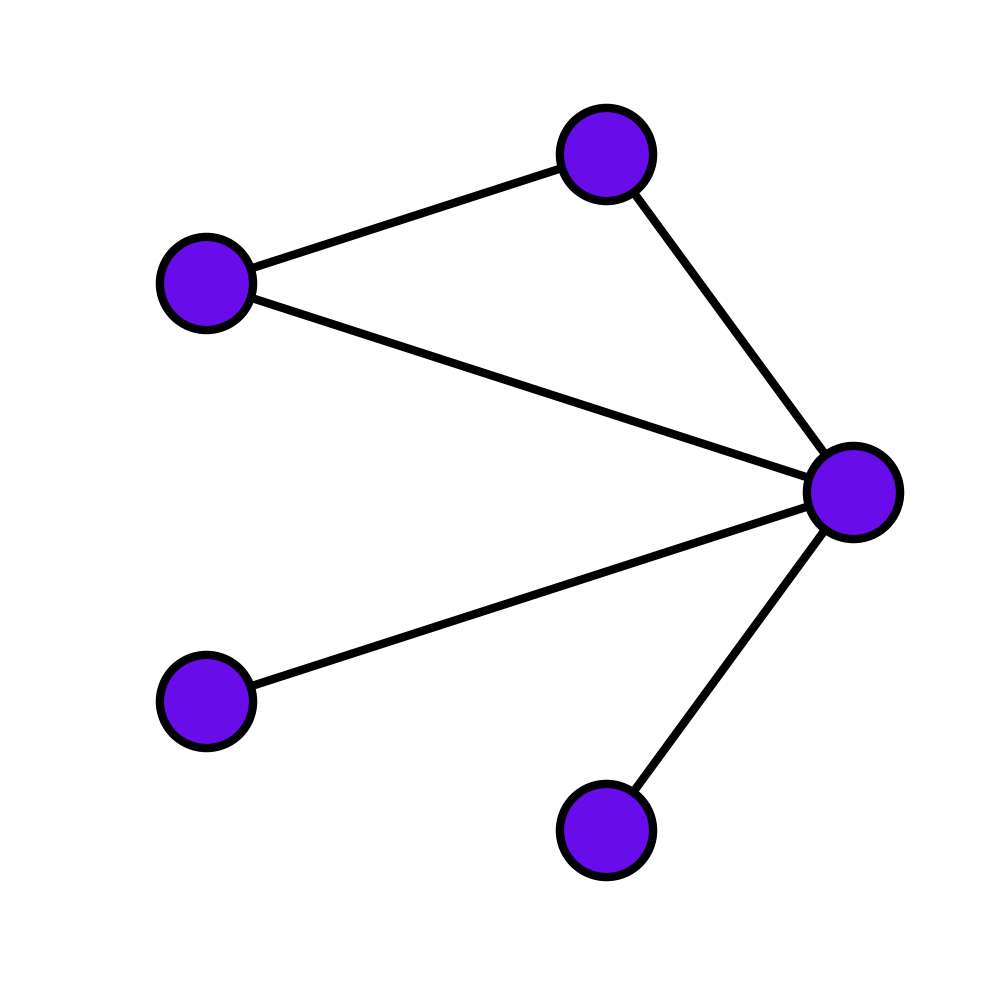} & \texttt{9915ff93} & 5.926E-5 & 1 & - & 19 & 12 & 32 & 6 & 31104 \\
  & \includegraphics[valign=c,width=\graphwidth]{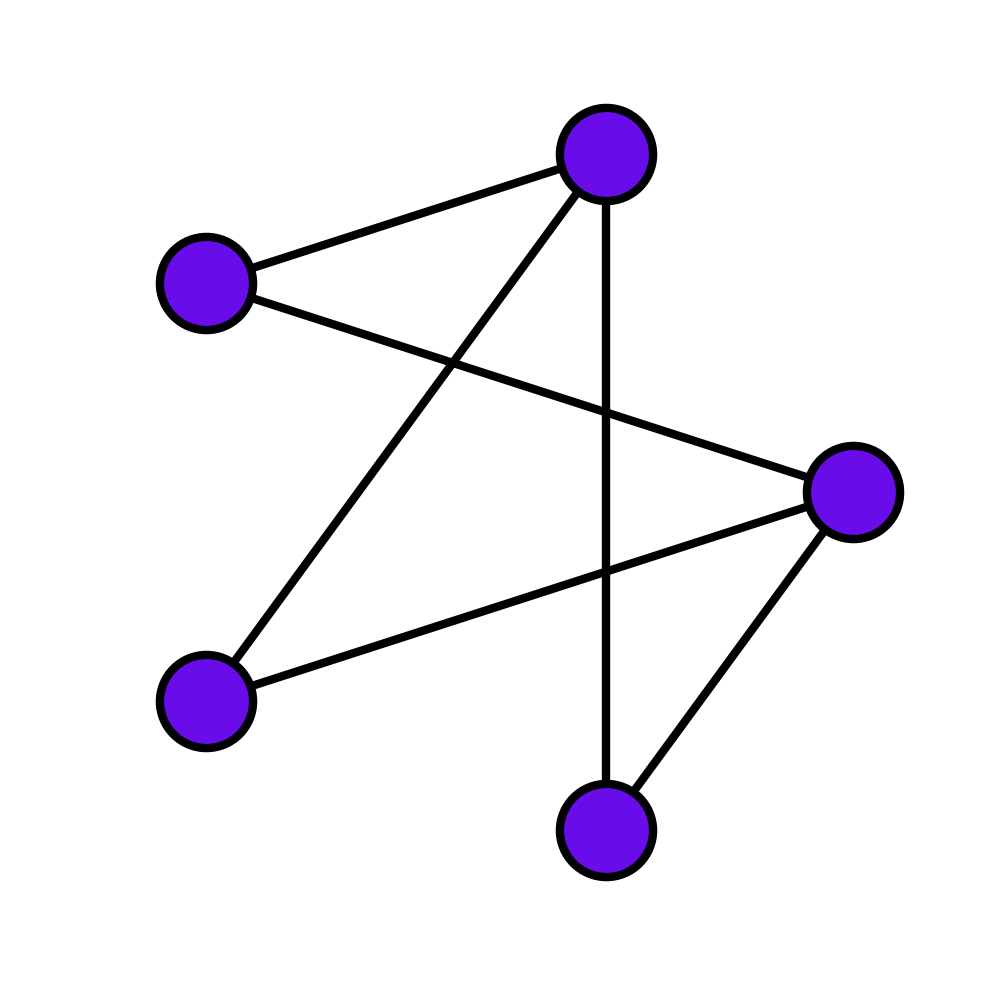} & \texttt{646c4ffd} & 1.194E-4 & 2 & - & 20 & 11 & 32 & 6 & 31104 \\
  & \includegraphics[valign=c,width=\graphwidth]{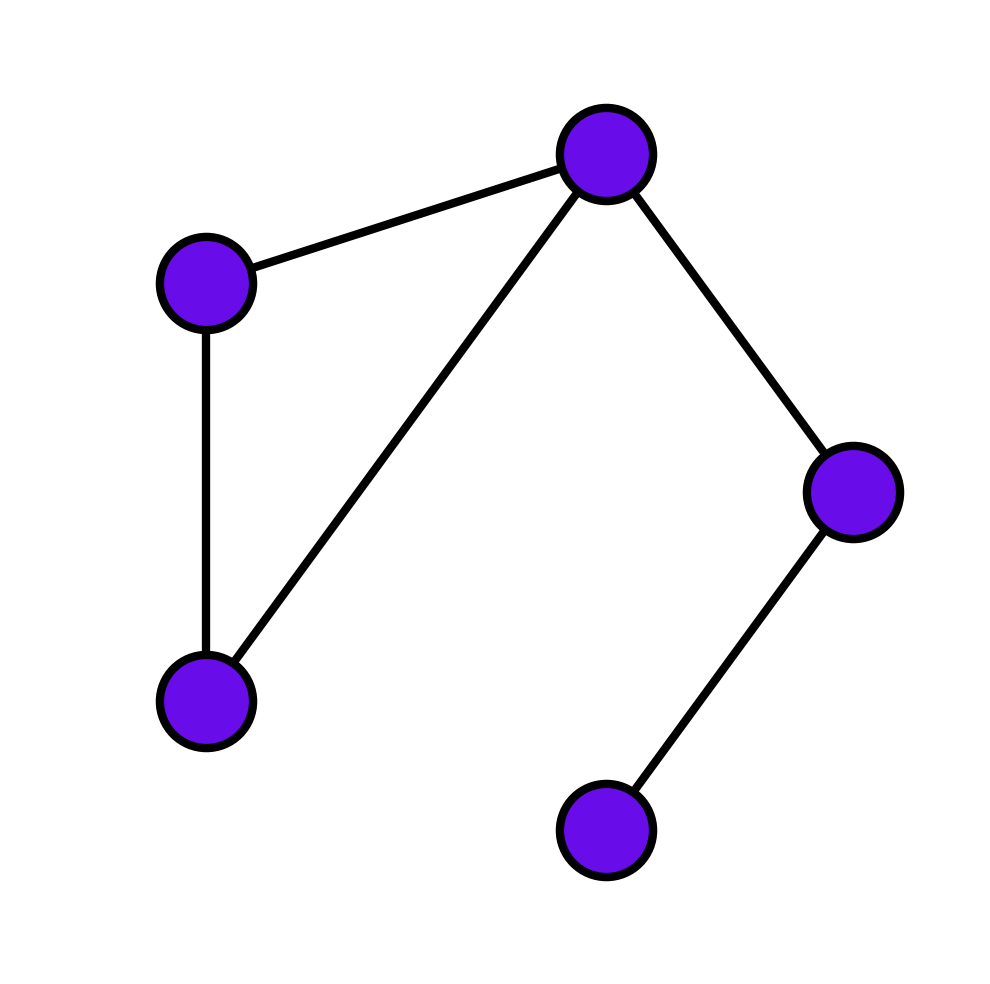} & \texttt{b2c25a19} & 3.464E-4 & 12 & - & 20 & 12 & 32 & 7 & 62208 \\
\bottomrule
\end{tabular}
\label{tab:discovered_circuits}
\end{table*}

In this section, we present circuits for graph states of increasing size obtained using our discovery pipeline. We consider connected graphs of size 3, 4, and 5 qubits, and compare our automatically discovered circuits against a fusion baseline approach. %
(Sec.~\ref{sec:5modeexample} presents a discovered circuit for Bell-state generation, which is local Clifford-equivalent to the single connected 2-node graph state). For each target graph state, we conducted parameter sweeps over the number of ancilla modes and total photon count. The maximum photon count per mode for the input state was set to 1 (i.e., the input state was $\kket{11 \ldots 1 0 \ldots 0}$). To address the non-convex optimization landscape with numerous local minima, we employed multiple random initializations and selected the best-performing solutions across different optimization runs. Additional implementation details are provided in Appendix~\ref{app:implementation}. Table~\ref{tab:discovered_circuits} presents our best results, selected based on success probability and minimal ancillary resource requirements. We consider a circuit to successfully prepare a target state if the fidelity between the prepared state and any member of the target state set $\mathcal{T}$ exceeds $0.9999$. The success probability is then the sum of all probabilities for conditional states exceeding this fidelity.

\subsubsection{3 Qubits}
For 3 qubits, there are 2 distinct connected graphs, both belonging to equivalence class 2. For the linear graph state, we discovered two distinct circuits. The first achieves $1.85\%$ success probability using 11 modes and 6 photons, with 3 ancilla measurement outcomes that herald target states. The second achieves higher success probability of $2.22\%$ using significantly more resources (15 modes and 10 photons) and produces 12 heralded target states. For the triangle graph state, we discovered a circuit with $3.125\%$ success probability using 12 modes and 6 photons, producing 8 heralded target states. These metrics match those of the analytically-derived $n=3$ GHZ circuit from Ref.~\cite{bartolucci_creation_2021}. 

\subsubsection{4 Qubits}
For 4 qubits, there are 6 distinct connected graphs grouped into 2 LC equivalence classes. Equivalence class 3 contains the star and complete graph states, which are local Clifford-equivalent to the 4-GHZ state. We discovered a star graph circuit achieving $0.46\%$ success probability using 14 modes and 9 photons, with 4 heralded target states. Our best complete graph circuit achieved $0.78\%$ success probability using 17 modes and 8 photons, producing target states for 24 different ancilla measurements. As a comparison point, the analytically-derived 4-GHZ circuit from Ref.~\cite{bartolucci_creation_2021} achieved $0.78\%$ success probability using 16 modes and 8 photons, with 16 heralded target states. Equivalence class 4 contains the remaining 4 graph states, including the linear and ring states. Our discovered circuits use 15--17 modes and 9 photons, achieving success probabilities ranging from $0.21\%$ to $0.36\%$. To our knowledge, these represent the first circuits reported for these states. The discovery of circuits for all 4-qubit graph states, including those with no previously known implementations, demonstrates the completeness of our search methodology.

\subsubsection{5 Qubits}
For 5 qubits, there are 21 distinct connected graphs grouped into 4 equivalence classes. Due to the computational complexity of 5-qubit optimizations and the large number of target states, we focused on discovering representative circuits from each equivalence class.

Equivalence class 5 contains the star and fully-complete graph states, which again are local Clifford-equivalent to the 5-GHZ state. Our best star graph circuit achieved $0.116\%$ success probability using 19 modes and 10 photons, producing target states for 12 different ancilla measurements. As a comparison point, the analytically-derived 5-GHZ circuit from Ref.~\cite{bartolucci_creation_2021} achieved $0.195\%$ success probability using 20 modes and 10 photons, with 32 heralded target states.

For equivalence class 6, we discovered two circuits: one achieving $0.0059\%$ success probability with 1 heralded target state, and another achieving $0.012\%$ success probability with 2 heralded target states. For equivalence class 7, we found a circuit achieving $0.035\%$ success probability that produces 12 target states. To the best of our knowledge, these represent the first known state preparation circuits for these particular 5-qubit graph states. Lastly, while computational limitations prevented exhaustive exploration of all 5-qubit states, our results span each equivalence class and establish new benchmarks for several graph families.

\subsection{Sparsified circuits}

\begin{figure*}[!htbp]
\includegraphics[width=1\textwidth]{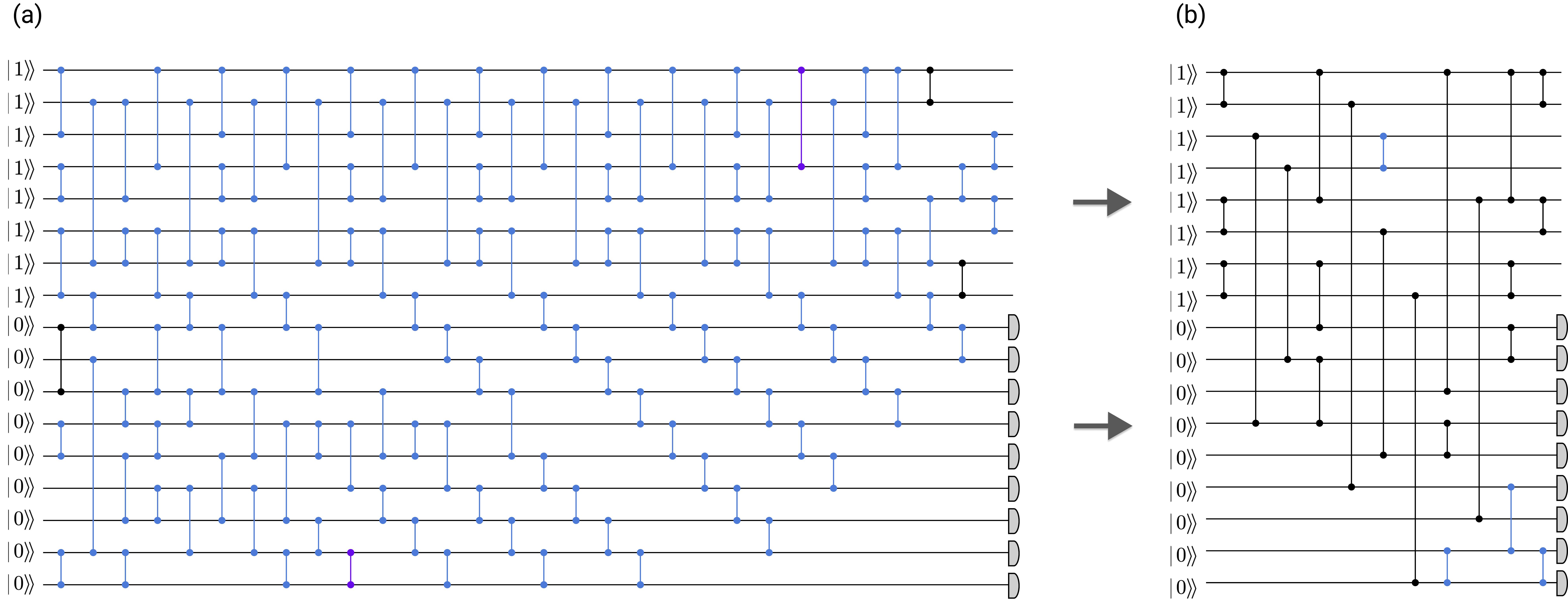}
    \caption{
    \label{fig:fully_connected_4q}
    Discovered photonic circuits for the fully connected 4-qubit graph state (complete graph $K_4$). (a) The compiled circuit produced by the Clements {\em et al.} decomposition applied to the transfer matrix returned by the first optimization pass, $U_1$. (b) The final, simplified circuit obtained after compiling the result of the second optimization pass. Following the conventions of Ref.~\cite{bartolucci_creation_2021}, each horizontal line represents a photonic mode. The initial Fock state is shown to the left, with modes 1-8 receiving an input photon. The circuit uses dual-rail encoding with modes 1-2, 3-4, 5-6, and 7-8 representing the four logical qubits. CZ-gate symbols denote beamsplitters that couple mode pairs. Black beamsplitters are standard 50:50 couplers ($R=1/2$), purple beamsplitters have $R=1/3$, and blue beamsplitters have arbitrary reflection coefficients. Note that the phaseshift angles $\phi_T$, $\phi_R$  are not specified here.
    } 
\end{figure*}

While our first optimization stage successfully identifies transfer matrices that prepare target graph states with high success probabilities, the resulting solutions typically require dense optical fabrics that would be challenging to experimentally implement on certain device architectures. The sparsification procedure developed in Sec.~\ref{sec:optimizationpass2} reduces this complexity without sacrificing performance. Across all discovered circuits, we observe beamsplitter count reductions of 50-80\% relative to fabric decompositions of generic unitaries, with many of the remaining beamsplitters exhibiting rational reflection coefficients. These simplifications suggest that optimal photonic circuits possess inherent mathematical structure that our regularization procedure successfully uncovers.

The second optimization stage begins with the optimized transfer matrix from the first pass, decomposes it using the Clements {\em et al.} fabric representation, and then removes unnecessary beamsplitters through regularized sparsification. As depicted in Fig.~\ref{fig:pipeline}~(d), the sparsified fabric is further compiled into a simplified non-local circuit representation. This two-step process---sparsification followed by compilation---often discovers circuits with substantially reduced complexity compared to a generic transfer matrix implementation. Fig.~\ref{fig:fully_connected_4q} illustrates the simplification afforded by this sparsification process for the complete 4-qubit graph state, comparing the initial circuit from the Clements {\em et al.} decomposition of transfer matrix $U_1$ (panel (a)) with the final sparsified circuit (panel (b)). Both circuits achieve identical $0.78\%$ success probability while preparing the desired target states. The sparsification procedure frequently yields circuits with rational reflection coefficients $R := \sin^2\theta$, where $\theta$ is the beamsplitter angle from Eq.~\ref{eq:beamsplitter}. In this particular circuit, 19 out of the 23 beamsplitters are 50:50 ($R=1/2$), indicating a strong preference for symmetric beam splitting. This prevalence of simple rational coefficients, combined with the dramatic reduction in circuit complexity, suggests that the optimal photonic circuits possess underlying mathematical structure that our optimization framework successfully uncovers.

\begin{figure*}[!htbp]
\includegraphics{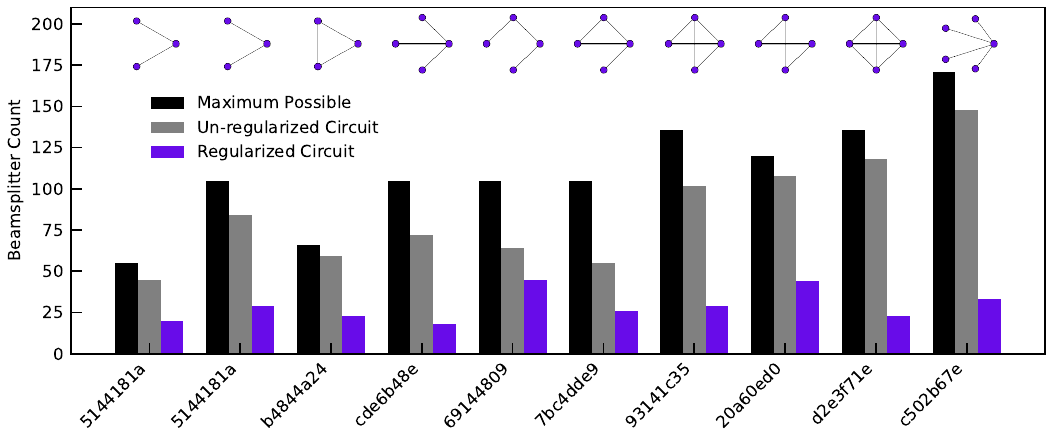}
    \caption{
    \label{fig:beamsplitter_reduction_plot}
    Optical circuit sparsification. Beamsplitter counts (excluding trivial and SWAP-equivalent operations) for the discovered circuits presented in Table~\ref{tab:discovered_circuits}. Solutions are labeled by the first 8 characters of the WL hash of the target graph state. For each solution, three counts are displayed: the theoretical maximum beamsplitter count for the Clements {\em et al.} decomposition of an $m$-mode unitary, $m(m-1)/2$ (black); the actual beamsplitter count from applying the Clements {\em et al.} decomposition to the first-pass optimized unitary $U_1$ (gray); and the final beamsplitter count after regularization and compilation (purple). Results are shown in the same order as in Table~\ref{tab:discovered_circuits}.
    } 
\end{figure*}

The circuit simplification procedure significantly reduces beamsplitter complexity, as quantified in Fig.~\ref{fig:beamsplitter_reduction_plot}. For each circuit presented in Table~\ref{tab:discovered_circuits}, we compare three beamsplitter counts: the theoretical maximum for a generic unitary matrix (black), the count from applying the Clements {\em et al.} decomposition to the stage-1 transfer matrix $U_1$ (gray), and the count in the final sparsified circuit (purple). The sparsification procedure achieves beamsplitter reductions of $\sim$50\% to $\sim$80\% relative to the initial decomposition.

The sparsification procedure often yielded benefits beyond circuit simplification, occasionally improving the underlying solution quality. We observed two primary mechanisms for these improvements. First, the regularization process sometimes identified unused modes that could be eliminated without affecting the heralded target states. For example, while the initial transfer matrix optimization for the 4-qubit star graph (\texttt{cde6b48e}) utilized a 15-mode ansatz, the subsequent sparsification revealed that only 14 modes were necessary to achieve the same performance. Notably, the Clements {\em et al.} decomposition of the first-stage result $U_1$ utilized all 15 modes, confirming that the mode reduction occurred as a result of the second-stage sparsification procedure. Second, the numerical rounding performed when replacing nearly-trivial or SWAP-equivalent beamsplitters with their exact equivalents occasionally improved circuit performance. This occurred because the optimization had converged to beamsplitter parameters that were close to, but not exactly, trivial values, and the rounding process corrected these small deviations to yield marginally better success probabilities. For instance, the success probability for the 4-qubit graph \texttt{93141c35} increased from 0.3545\% to 0.3579\% following sparsification. Sparsification results are not reported for circuits \texttt{9915ff93}, \texttt{646c4ffd}, and \texttt{b2c25a19} due to insufficient beamsplitter reduction during the regularization procedure. 

\subsection{Comparison to fusion-based graph state generation}

\begin{figure*}[!htbp]
\includegraphics{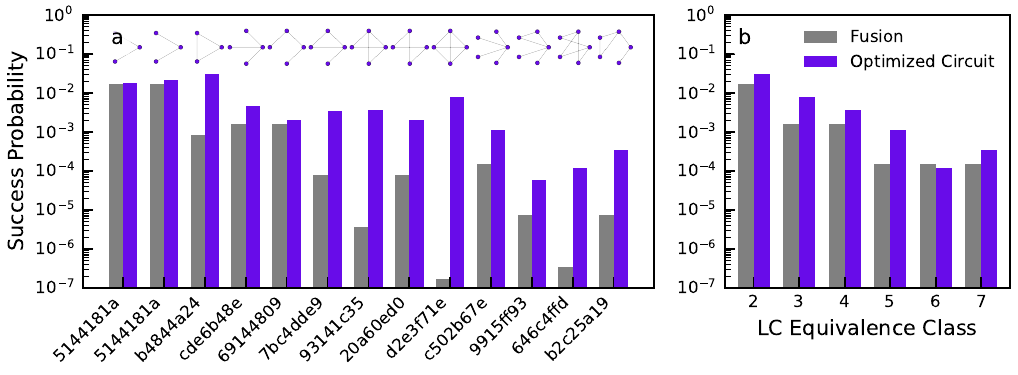}
    \caption{
    \label{fig:fusion_comparison_plot}
    Comparison of success probabilities: optimized circuits vs.\ fusion baseline.
    (a) Success probabilities for each optimized circuit in Table~\ref{tab:discovered_circuits} (purple) compared against the Type-I fusion baseline (gray), which constructs the target graph state by fusing Bell pairs according to Eq.~\eqref{eq:fusion}. Circuits are ordered as in Table~\ref{tab:discovered_circuits}.
    (b) Data from (a) aggregated by local complementation (LC) equivalence class, showing the maximum success probability achieved within each class. For fusion, this corresponds to the minimal-edge representative; for optimized circuits, the best-performing circuit discovered.
    } 
\end{figure*}

We compare our numerically discovered graph state preparation circuits against a fusion baseline approach that constructs graph states by fusing Bell pairs. To ensure a meaningful comparison under similar resource constraints, we consider a non-multiplexed fusion scheme without recycling or ancillary enhancement. While boosted fusion schemes using ancillary photonic states can achieve higher success probabilities, and multiplexed approaches are standard in practice and can achieve arbitrarily high success probabilities through repeated attempts, both require substantially more resources (additional photons, complex switching networks). We deliberately exclude such enhancements in order to isolate the advantages of circuit structure and resource allocation, rather than repeated sampling or ancillary overhead. This comparison isolates the benefits of direct synthesis under fixed resource budgets, complementing other analyses that might emphasize different trade-offs.

The fusion scheme begins by preparing one Bell pair for each edge in the graph. Each Bell pair is generated probabilistically and independently with success probability $p_{\text{Bell}}$. Fusion operations join Bell pairs according to the graph topology. The total number of required fusion operations is $2E - N$, where $E$ is the number of edges and $N$ is the number of vertices. Since the initial state comprises $2E$ photons and the final graph state comprises $N$ photons, each fusion operation must consume exactly one photon. Consequently, this scheme necessitates the use of Type-I fusion gates, as Type-II gates would consume two photons and fail to generate the target state. The overall success probability for constructing the graph state in a single attempt is therefore:
\begin{equation}
    \label{eq:fusion}
    p_{\text{success}} = p_{\text{Bell}}{}^{E} \, p_{\text{fusion}}{}^{2E - N} \,,
\end{equation}
which scales exponentially poorly with the number of edges. In contrast, our direct synthesis approach optimizes the entire circuit holistically, avoiding this composition penalty.

We compare against a fusion baseline characterized by two parameters: Bell pair generation probability $p_{\text{Bell}} = 3/16$ and fusion success probability $p_{\text{fusion}} = 1/2$. For Bell pair generation using four photons, the optimal success probability is $3/16 \approx 18.75\%$~\cite{bartolucci_fusion-based_2023}, which can be increased to $1/4$ through distillation or additional photon schemes that we exclude to maintain resource parity. The fusion success probability $p_{\text{fusion}} = 1/2$ is characteristic of canonical unboosted Type-I fusion~\cite{browne_resource-efficient_2005}. Without multiplexing, any failure in Bell pair preparation or fusion requires restarting the entire process.

Figure~\ref{fig:fusion_comparison_plot} compares our optimized circuits against the Type-I fusion baseline. In Panel (a), we compare success probabilities for the direct generation of each target graph state. Our optimized circuits outperform the fusion baseline in all cases, often by several orders of magnitude. However, this comparison is incomplete, as both methods can be optimized by exploiting LC equivalence. For the fusion baseline, one can generate the minimal-edge representative of the target LC class and subsequently transform it into the desired state via local Clifford operations. This approach is advantageous because the fusion success probability scales inversely with the number of edges: each additional edge introduces a penalty factor of $p_{\text{Bell}} \times p_{\text{fusion}}^2 = 3/64$. For our optimized circuit method, as explained above, a circuit that prepares a graph state $\ket{G}$ under Pauli equivalence may be converted to a circuit that prepares a LC-equivalent graph state $\ket{G'}$ through the application of a local Clifford. Given that our numerical optimizations exploit Pauli equivalence, it is more appropriate to compare the methods by LC equivalence class rather than by specific graph topology. This comparison is depicted in Panel (b). Our circuits surpass the LC-optimized fusion baseline in LC classes 2, 3, 4, 5, and 7, with substantial improvements in several cases. For example, in LC class 3 (4Q), the fully connected graph circuit achieves $7.813 \times 10^{-3}$ compared to $1.648 \times 10^{-3}$ for fusion---a $4.7\times$ improvement. In LC class 5 (5Q), the star graph circuit achieves $1.157 \times 10^{-3}$ versus $1.545 \times 10^{-4}$, a $7.5\times$ improvement.

These results validate our central thesis: gradient-based optimization, when properly structured with domain-specific constraints and regularization, can discover photonic circuits that match or exceed human-designed solutions. The progression from abstract transfer matrices to sparsified, hardware-ready circuits demonstrates that automated discovery need not sacrifice experimental feasibility for performance. We now discuss the broader implications of these findings and identify directions for extending this framework.

\section{Discussion \label{sec:discussion}}

Our work advances automated photonic design by demonstrating that a gradient-driven discovery pipeline can deliver resource-state generators (RSGs) with rational reflection coefficients suitable for experimental implementation. An FFT-based strong-simulation subroutine, operating within a bipartite-subspace optimization, enables exploration of linear optical transfer matrices for circuits with up to 20 spatial modes and 12 photons. This approach discovered the first known RSG circuits for several target graph states. The bipartite subspace restriction reduces the exponent of the time complexity for our FFT-based statevector simulation algorithm from $m$ to $m/2$ for an $m$-mode circuit in the symmetric case where half the modes are ancillas. This exponential speedup makes 5-qubit discoveries feasible on commodity hardware.

The pipeline's second, regularization pass then converts dense transfer matrices into hardware-friendly fabrics. For example, we discovered a 17-mode unitary that prepares the 4-qubit complete graph state. While a generic 17-mode unitary would require 136 beamsplitters and the optimized transfer matrix from the first stage required 122, the final, sparsified circuit uses only 23 beamsplitters---an 81\% reduction. Smaller optical depth translates directly into fewer loss channels and a simpler specification set for fabrication. 

The sparsification procedure primarily aims to reduce optical depth, but unexpectedly reveals underlying mathematical structure. The remaining beamsplitters after sparsification frequently exhibit simple rational reflection coefficients. For the complete 4-qubit graph state circuit, 19 of the 23 remaining beamsplitters are 50:50 couplers ($R = 1/2$), while the remaining four also appear to have rational coefficients. We attribute this phenomenon to the optimization landscape structure: the initial stage seeks any viable solution regardless of symmetry, while sparsification deforms this solution toward the simplest representation that preserves the desired target state preparation.

However, not all circuits achieved significant simplification through this procedure. Sparsification results are not reported for the 5-qubit graph state circuits with graph hash ids \texttt{9915ff93}, \texttt{646c4ffd}, and \texttt{b2c25a19} due to insufficient beamsplitter reduction. This limitation stems from both computational constraints---fewer optimization runs were feasible due to longer iteration times with higher photon and mode counts---and the apparent correlation between sparsification efficacy and initial solution quality. We speculate that circuits exhibiting relatively low success probabilities lack the underlying mathematical structure that facilitates effective sparsification, consistent with our observation that high-performance solutions more readily yield compact representations. This finding has practical implications for circuit discovery: optimization efforts should prioritize achieving high success probabilities in the initial transfer matrix stage, as these solutions are more likely to yield implementable, hardware-efficient circuits after sparsification.

Ballistic synthesis has the potential to outperform fusion-based graph state generation whenever switching resources are scarce. 
In local-complementation class 3 (containing the four qubit star graph and fully connected graph), our best circuit achieves a success probability of $7.813\times10^{-3}$ versus $1.648\times10^{-3}$ for the fusion-based baseline (a $4.7\times$ advantage) (Table I, row \texttt{d2e3f71e}). 
The 5-qubit star graph state generator in class 5 exhibits an even larger margin, $1.157\times10^{-3}$ compared with $1.545\times10^{-4}$ (a $7.5\times$ improvement) (Table I, row \texttt{c502b67e}). Taken together, the circuits reported here constitute a menu of high-yield RSGs that can seed arbitrarily large, fault-tolerant graph states. The reduced beamsplitter count and rationalized reflection coefficients make them well-suited to lithographic or femtosecond-laser writing processes, accelerating experimental timelines.

The pipeline also corroborates previously posed analytic performance bounds. The fully connected 3 and 4 qubit graph states are locally equivalent to a 3-GHZ and 4-GHZ states. In these cases the optimizer rediscovered the Bartolucci {\em et al.} success frontier---3.125\% for 3 qubits, and 0.7813\% for 4 qubits. This numerical evidence supports the conjecture that the Bartolucci result is indeed optimal within the ballistic, dual-rail model.

Notably, our discovered circuits yield different success probabilities even for locally equivalent graph states. For instance, graphs \texttt{69144809} and \texttt{7bc4dde9} both belong to LC class 4, meaning their corresponding graph states are related by local Clifford operations. Despite using identical resources (15 modes, 9 photons), our best circuits for these states achieved success probabilities of 0.205\% and 0.347\%, respectively. As discussed in Sec.~\ref{sec:graphstateresults}, exploiting Pauli equivalence allows a circuit generating a graph state $\ket{G}$ to be converted into one generating an LC-equivalent state $\ket{G'}$ without altering the success probability. The observed variation must therefore reflect differences in the optimization landscape: gradient-based methods converge to different local minima depending on the specific target graph state within an LC class, as well as on initialization and optimization hyperparameters. This interpretation is supported by our LC class 5 results, which fall short of known analytical benchmarks: our 5-qubit star graph circuit achieves only 0.116\% success probability, compared to 0.195\% for the analytical 5-GHZ circuit of Ref.~\cite{bartolucci_creation_2021}. Since the 5-GHZ state is LC-equivalent to class 5 graph states and thus lies within our theoretical search space, this gap confirms that our optimization has not yet located the global optimum. Further confirmation for this conclusion is provided by direct optimization for target GHZ states for $n=4, 5$: using 100 runs with fixed hyperparameters, the optimizer failed to converge to successful solutions, substantiating the hypothesis the performance gap arises from differences in the optimization landscape rather than the expressivity of the ansatz. These findings motivate future work on alternative optimization strategies and systematic searches over target representations within equivalence classes.

Several directions would advance this work toward practical implementation and broader applicability. Hardware validation through implementation of discovered circuits on photonic processors would validate simulation predictions and assess real-world performance. This validation effort should be coupled with extending the simulation framework to include hardware noise models, loss, and imperfections for robust circuit designs that account for experimental realities. Computational scaling represents another priority: leveraging distributed GPUs and memory management techniques could scale beyond current 5-qubit limitations to reach the larger system sizes required for practical quantum applications. Finally, exploring broader equivalence classes is a natural next step. Extensions accommodating the single-qubit Clifford class or the single-qubit continuous-unitary class with adaptive feedforward could expand the accessible target space and improve success rates. In addition, mode permutations define a distinct equivalence relation that breaks the dual-rail encoding; accounting for it would extend the scope of our method.

\section{Conclusion \label{sec:conclusion}}
We have introduced a two-stage, differentiable design framework that autonomously discovers heralded photonic circuits for generating graph states up to five qubits. Leveraging an FFT-based strong-simulation algorithm restricted to a single rail-photon sector (a fixed split between signal and ancilla)---which halves the exponential scaling relative to a full-state FFT---together with a novel sparsification routine, we produced circuits that (i) operate on up to twenty modes and twelve photons, (ii) cut optical depth by more than 80\% relative to the unsparsified solutions, and (iii) raise ballistic success probabilities by factors of 4.7 (4-qubit) and 7.5 (5-qubit) over fusion baselines. For several graph topologies, these represent the first known state preparation circuits. These results supply a ready-to-fabricate library of high-yield RSGs and mark an essential step toward scalable, resource-efficient fusion-based quantum computing.

\subsection*{Acknowledgments}
We are grateful to all other colleagues at Q-CTRL whose technical, product engineering, and design work has supported the results presented in this paper. We thank Viktor Perunicic for discussions at an early stage of this project. We also thank Terry Rudolph and Patrick Birchall for helpful discussions on GHZ and graph state generator comparisons, as well as benchmarking considerations for numerical photonic packages. Lastly, we thank an anonymous referee for insightful comments that clarified our interpretation of results across LC equivalence classes as well as our characterization of the fusion baseline.

\appendix 

\section{Implementation details \label{app:implementation}}

Here, we provide technical details for our optimization procedure. Both optimization stages (described in Sec.~\ref{sec:optimizationpass1} and Sec.~\ref{sec:optimizationpass2}, respectively) employed first-order gradient-based optimization using the Adam optimizer \cite{kingma_adam_2014}. We managed the optimization workflow using \textit{PyTorch Lightning}, which provided automated early stopping, regularization, and learning rate scheduling capabilities. Both stages optimization used a \textsc{OneCycleLR} schedule \cite{smith_super-convergence_2019}, while the sparsification stage additionally employed a \textsc{CosineAnnealingLR} schedule \cite{loshchilov_sgdr_2016} to control the strength of the regularization term. Optimization continued until the loss plateaued or reached a maximum of 1000 iterations, whichever occurred first.

All computations were performed on Nvidia RTX A5000 GPUs with 24 GB memory capacity, using single-precision complex arithmetic (i.e., using the \texttt{torch.complex64} data type). This precision choice balanced numerical accuracy with computational efficiency, particularly important given the exponential scaling of our tensor representations with system size. 

\section{FFT-based strong-simulation algorithm: additional details \label{app:FFTdetails}}
This appendix details our FFT-based strong-simulation method for linear-optical polynomials. We review polynomial multiplication via FFT, then derive the photon-number sector decomposition underlying \textsc{BipartitePolyMulFFT} and analyze its complexity. After illustrating both algorithms on the 5-mode Bell-state example, we compare our approach to \textsc{SLOS}, discussing the trade-off between asymptotic efficiency and compatibility with automatic differentiation.

\subsection{Polynomial multiplication via the Fourier transform}
We briefly review how multivariate polynomial multiplication may be achieved via the Discrete Fourier Transform (DFT), treating first univariate polynomials and then extending to the multivariate case. Consider two degree-$n$ polynomials with coefficients in $\mathbb{C}$: $f(z) = \sum_{i=0}^n f_i \, z^i$ and $g(z) = \sum_{i=0}^n g_i \, z^i$, and denote the coefficient sequences as $\bm{f} = (f_0, f_1, ..., f_n)$ and similarly, $\bm{g} = (g_0, g_1, ..., g_n)$. The product 
\begin{equation}
	h(z)=f(z)g(z)=\sum_{k=0}^{2n} h_k z^k
\end{equation}
has coefficients given by the Cauchy product (linear convolution):
\begin{equation}
	h_k=\sum_{i=0}^{n} f_i\,g_{k-i}\quad (0\le k\le 2n) \,,
\end{equation}
which we write more compactly as $\bm{h} = \bm{f} * \bm{g}$, and where we interpret $g_{k-i}=0$ whenever $k-i \notin \{0, 1, ..., n\}$.

To compute this via the DFT, choose a transform
length $N\ge 2n+1$ and form length-$N$ sequences by zero-padding the
coefficients: ${f_{i} = g_{i} = 0}$ for $i \in \{n+1, ..., N-1\}$. By the convolution theorem,
\begin{equation}
	\bm{f} * \bm{g}  = \mathrm{DFT}_N^{-1} \left( \mathrm{DFT}_N( \bm{f} )\,\odot\,\mathrm{DFT}_N( \bm{g} ) \right) \,.
\end{equation}
Here $\odot$ denotes the elementwise product. Zero-padding to $N \ge 2n+1$ ensures the circular convolution computed by the DFT (i.e., convolution mod $N$) coincides with the desired linear convolution on the first $2n+1$ entries.

Next, we consider the extension to multivariate polynomials (defined over $\mathbb{C}^m$). Let
\begin{equation}
	f(z) = \sum_{i_1=0}^n \ldots \sum_{i_m=0}^n f_{i_1, \ldots, i_m} \, z_1{}^{i_1} \ldots z_m{}^{i_m} \,,
\end{equation}
and similarly for $g$. Now, $\bm{f}, \bm{g}$ denote the $m$-dimensional coefficient tensors. The product polynomial $h$ has degree at most $2n$ in each variable. The coefficients are given by
\begin{equation}
	h_{k_1,\ldots,k_m} =\!\!\sum_{i_1=0}^{n}\cdots\sum_{i_m=0}^{n} f_{i_1,\ldots,i_m} \, g_{k_1-i_1,\ldots,k_m-i_m} \,, 
\end{equation}
where $0\le k_\ell\le 2n$ and where we again adopt the convention that $g_{k_1-i_1, \ldots, k_m - i_m}$ is zero whenever any index is less than zero or greater than $n$. In tensor notation,
\begin{equation}
	\bm{h}=\bm{f} * \bm{g},
\end{equation}
where now $*$ denotes the $m$-dimensional linear convolution. As before, we pad each dimension with zeros to ensure that the circular convolution computed by the DFT matches the linear convolution. 

For the linear-optical setting considered in this work, the relevant polynomial is generated by a unitary $U$ acting on the mode creation operators of an input Fock state $\kket{\bm{n}}$. The output polynomial coefficients can then be obtained using an FFT-based implementation of this multivariate convolution, as explained in the main text. Algorithm~\ref{alg:polymulfft} gives pseudocode for this procedure, which we call \textsc{PolyMulFFT}. Note that the requirement that $U$ be unitary is physical, not mathematical, and the multiplication algorithm is valid even when $U$ is a general complex-valued matrix. 

\begin{algorithm}[H]
\caption{PolyMulFFT}\label{alg:polymulfft}
\begin{algorithmic}
\Require
\Statex Transfer matrix $U$
\Statex Input occupations $\mathbf{n}=(n_1,\ldots,n_m)$ with $\sum_i n_i=n$
\Ensure Output tensor $T$ of shape $(n{+}1)^m$ with entries $T_{\mathbf{n}}$
\State $S \gets (n{+}1,\ldots,n{+}1)$ \Comment{Set tensor shape}
\State $G \gets \mathbf{1}_S$ \Comment{Initialize tensor to all-1's}
\For{$i=1,\ldots,m$}
  \State $T^{(i)} \gets \mathbf{0}_S$ \Comment{Initialize tensor to all-0's}
  \For{$j=1,\ldots,m$}
    \State $T^{(i)}[\mathbf{e}_j] \mathrel{+}= U_{ji}$ \Comment{$\mathbf{e}_j$ is 1 in mode $j$, 0 elsewhere}
  \EndFor
  \State $F^{(i)} \gets \mathrm{FFT}(T^{(i)})$ \Comment{$m$-dim FFT on shape $S$}
  \State $G \gets G \odot \big(F^{(i)}\big)^{\,n_i}$ \Comment{Power and pointwise mult.}
\EndFor
\State $T \gets \mathrm{FFT}^{-1}(G)$
\State $T \gets T\ /\ \sqrt{\prod_{i=1}^m n_i!}$ \Comment{Fock state normalization}
\State \Return $T$
\end{algorithmic}
\end{algorithm}

\subsection{Bipartite polynomial multiplication algorithm \label{app:bipartite_algorithm}}
Next, we derive Eq.~\ref{eq:polynomial_sector_decomposition}, the decomposition of the state polynomial into sectors based on the number of photons. This facilitates the development of \textsc{BipartitePolyMulFFT}, the restricted variant of the strong statevector simulation algorithm where only a single sector is computed.

Recall that the LOQC evolution rule, Eq.~\ref{eq:LOQCtransformation}, is
\begin{equation}
    a^{\dagger}_{\text{in}, i} = \sum_{j=1}^m a^{\dagger}_{\text{out}, j} U_{ji} \,.
\end{equation}
(We use subscripts to explicitly indicate the distinction between input and output operators in what follows to avoid ambiguity.) It will be useful to introduce the signal and ancilla components of each transformed input operator:
\begin{equation}
	a_{\text{in}, S, i}^{\dagger} := \sum_{j=1}^{m_S} a_{\text{out}, j}^{\dagger} U_{j i} \, \qquad a_{\text{in}, A, i}^{\dagger} := \sum_{j=m_S + 1}^{m} a_{\text{out}, j}^{\dagger} U_{j i}
\end{equation}
The output state polynomial may be written as
\begin{equation}
	\label{eq:prod_form}
	P( \bm{a}_{\text{out}}^{\dagger}) = \prod_{i=1}^m \left( a_{\text{in}, S, i}^{\dagger} + a_{\text{in}, A, i}^{\dagger}  \right)^{n_i} \,,
\end{equation}
where the $n_i$ are determined by the input state, and $\bm{a}_{\text{out}}^{\dagger}$ denotes the full set of output creation operators. Each term may be expanded using the binomial theorem:
\begin{equation}
	 \left( a_{\text{in}, S, i}^{\dagger} + a_{\text{in}, A, i}^{\dagger}  \right)^{n_i} = \sum_{s_i = 0}^{n_i} \binom{n_i}{s_i} \left( a_{\text{in}, S,i}^{\dagger} \right)^{s_i} \left( a_{\text{in}, A, i}^{\dagger} \right)^{n_i - s_i} \,,
\end{equation}
where $s_i$ may be interpreted as the number of photons sent from input mode $i$ to the signal output modes; the remaining $n_i - s_i$ are sent to the ancilla output modes. The factor $\binom{n_i}{s_i}$ is the binomial weight for choosing which $s_i$ of the $n_i$ photons go to the signal output modes.

Inserting this into Eq.~\ref{eq:prod_form} yields
\begin{equation}
	P( \bm{a}_{\text{out}}^{\dagger}) = \sum_{s_1, s_2, \ldots, s_m} \prod_{i=1}^m \binom{n_i}{s_i} \prod_{i=1}^m \left( a_{\text{in}, S,i}^{\dagger} \right)^{s_i} \prod_{i=1}^m \left( a_{\text{in}, A,i}^{\dagger} \right)^{n_i - s_i} \,,
\end{equation}
where the sum runs over integers $s_i \in \{0,\ldots,n_i\}$. Next, organize the sum according to the total number of signal photons, $n_S = \sum_{i=1}^m s_i$.
This leads to the decomposition
\begin{equation}
	P( \bm{a}_{\text{out}}^{\dagger}) =  \sum_{n_S=0}^{n} P^{(n_S,\,n-n_S)}(\bm{a}_{\text{out}}^\dagger) \,,
\end{equation}
where
\begin{widetext}
\begin{equation}
	\label{eq:sector_sum_derivation}
	P^{(n_S,\,n-n_S)}(\bm{a}_{\text{out}}^\dagger) = \sum_{\substack{s_1+\cdots+s_m= n_S \\ 0 \le s_i \le n_i}} \prod_{i=1}^m \binom{n_i}{s_i} \prod_{i=1}^m \left( a_{\text{in}, S,i}^{\dagger} \right)^{s_i} \prod_{i=1}^m \left( a_{\text{in}, A,i}^{\dagger} \right)^{n_i - s_i} \,,
\end{equation}
\end{widetext}
and where each component $P^{(n_S,\,n-n_S)}$ is homogeneous of degree $n_S$ in the signal output creation operators and degree ${n_A = n-n_S}$ in the ancilla output creation operators. This decomposition enables the separate computation of the output states for each sector, where only amplitudes corresponding to a specific signal-ancilla photon split $(n_S, n_A)$ are evaluated. 

We call the resulting algorithm, summarized in Algorithm~\ref{alg:bipartite_algorithm}, {\textsc{BipartitePolyMulFFT}}, reflecting the factorization of each sector's Fock space into signal and ancilla subspaces: $\Phi_{m_S, n_S} \times \Phi_{m_A, n_A}$. Photon-number conservation is exploited to convert each dense polynomial to a sparse fixed-sum representation before the final contraction. Since almost all entries of the dense tensors vanish, this sparse sector indexing substantially reduces time and memory overhead.
The conversion to a sparse format is performed by indexing/deindexing subroutines \textsc{IndexToFockState} and \textsc{ModeSplitToIndex}.
\textsc{IndexToFockState}$(m, n)$ returns a map, $\mathcal I$, associating indices $\{1,2,\cdots,|\mathcal I|\}$ with tuples that represent Fock states of $n$ photons distributed among $m$ modes. $\mathcal I$ can also be interpreted as a vector whose elements are tuples of length $m$, used to index into a dimension-$m$ tensor and produce a vector that is a sparse representation. $\mathcal S = \textsc{ModeSplitToIndex}(\mathbf n, n_S)$ maps tuples $\mathbf s$ representing solutions of $\sum_i s_i = n_S,\ 0\leq s_i\leq n_i$ to indices $\{1,2,\cdots,|\mathcal S|\}$. The final step of the algorithm constructs a bipartite state from the sparse representation.

\begin{algorithm}[H]
\caption{BipartitePolyMulFFT}\label{alg:bipartite_algorithm}
\begin{algorithmic}
\Require 
\Statex Transfer matrix $U$
\Statex Input occupations $\mathbf{n}$ with $\sum_i n_i=n$
\Statex Rail mode split $m_S{+}m_A{=}m$
\Statex Target photon split $n_S{+}n_A{=}n$
\Ensure Sparse bipartite state on sector $(n_S, n_A)$
\State $U_S \gets U\left[1{:}m_S,1{:}m\right],\quad U_A \gets U\left[m_S{+}1{:}m,1{:}m\right]$
\State $\mathcal I_S \gets \textsc{IndexToFockState}(m_S, n_S)$
\State $\mathcal I_A \gets \textsc{IndexToFockState}(m_A, n_A)$
\State $\mathcal S \gets \textsc{ModeSplitToIndex}(\mathbf n, n_S)$
\State $M_S \gets \textsc{Empty}(|\mathcal I_S|, |\mathcal S|)$
\State $M_A \gets \textsc{Empty}(|\mathcal S|, |\mathcal I_A|)$
\For{each per-mode split $\mathbf s$ with $\sum_i s_i = n_S$ and $0\le s_i\le n_i$}
  \State $\mathbf a \gets \mathbf n - \mathbf s$
  \State $c \gets \prod_{i=1}^m \binom{n_i}{s_i}$ \Comment{Combinatorial factor}
  \State $\mathsf P_S \gets \textsc{PolyMulFFT}(U_S, \mathbf s)$
  \State $\mathsf P_A \gets \textsc{PolyMulFFT}(U_A, \mathbf a)$
  \State $M_S\left[1{:}|\mathcal I_S|, \mathcal S(\mathbf s)\right] \gets c\ \mathsf P_S\left[\mathcal I_S\right]$ \Comment{Tensor sparsification}
  \State $M_A\left[\mathcal S(\mathbf s), 1{:}|\mathcal I_A|\right] \gets \mathsf P_A\left[\mathcal I_A\right]$
\EndFor
\State $B\gets M_S M_A$ \Comment{Shape $(|\mathcal I_S|,|\mathcal I_A|)$}
\State \Return $\sum_{i=1}^{|\mathcal I_S|} \sum_{j=1}^{|\mathcal I_A|} B[i, j] \mathcal I_S(i) \otimes \mathcal I_A(j)$
\end{algorithmic}
\end{algorithm}

We now analyze the time and space complexity of \textsc{BipartitePolyMulFFT}. A key quantity appearing in the analysis is the number of admissible per-mode splits iterated over in the main for-loop of the algorithm:
\begin{equation}
  |\mathcal{S}| \;:=\; \#\Big\{(s_1,\ldots,s_m)\in\mathbb{Z}_{\ge 0}^m \,\Big|\, \sum_{i=1}^m s_i = n_S,\; 0\le s_i \le n_i \Big\} \,.
\end{equation}
A general closed-form expression for $|\mathcal{S}|$ is not available; however, when each input mode contains at most one photon, $n_i \in \{0, 1\}$, then $|\mathcal{S}| = \binom{n}{n_S}$.

The contributions to the space complexity are: the intermediate dense tensor representations of the polynomial for each mode type (signal and ancilla), $\mathsf{P}_S$ and $\mathsf{P}_A$, the sparse tensors $M_S, M_A$, and the final bipartite contraction $B$. The peak space complexity is therefore
\begin{align}
	\label{eq:space_bipartite}
	\mathrm{Space} &= O\Big(\max\big\{ (n_S{+}1)^{m_S},\ (n_A{+}1)^{m_A}, \\
	&\qquad\qquad |\mathcal{S}| \,|\Phi_{m_S,n_S}|,\  |\mathcal{S}| \,|\Phi_{m_A,n_A}|, \nonumber \\ 
	&\qquad\qquad |\Phi_{m_S,n_S}| \, |\Phi_{m_A,n_A}| \big\}\Big). \nonumber
\end{align}

The dominant time costs are: (i) for each admissible split, two applications of \textsc{PolyMulFFT}, and (ii) the final contraction over the partition index (which contributes a factor of $|\mathcal{S}|$). A single $m$-dimensional FFT on a tensor of size $(n + 1)^m$ costs $O \big(m (n+1)^{m}\log(n + 1)\big)$; the computation of $\mathsf{P}_S$ and $\mathsf{P}_A$ each require $O(m)$ such FFTs per split (one per input-mode factor in the DFT-convolution product). Hence the total runtime is
\begin{align}
	\label{eq:time_bipartite}
	\mathrm{Time} &= O\Big(\max\big\{|\mathcal{S}| \big(m_S^2 \,(n_S + 1)^{m_S}\log(n_S + 1) \\
	& \qquad \qquad + m_A^2 \, (n_A + 1)^{m_A}\log(n_A + 1)\big) \,, \nonumber\\
	& \qquad \qquad |\mathcal{S}| \, |\Phi_{m_S, n_S}|\,|\Phi_{m_A, n_A}|\big\}\Big) \,. \nonumber
\end{align}

To identify the dominant terms, we note that binomial coefficients satisfy $\binom{n}{k} \leq 2^n$, with the maximum value (achieved at $k \approx n/2$) growing as ${\binom{n}{n/2} \sim 2^n/\sqrt{\pi n/2}}$. More generally, for the Fock space dimension ${|\Phi_{m,n}| = \binom{m+n-1}{n}}$, Stirling's approximation yields growth no faster than $c^n/\sqrt{n}$ for some constant $c$. In contrast, the dense tensor dimensions $(n_S+1)^{m_S}$ and $(n_A+1)^{m_A}$ exhibit exponential growth with bases that scale with $n$. Therefore, for sufficiently large $n$, the space complexity is dominated by the dense representations for the signal and ancilla polynomials, corresponding to $\mathsf{P}_S$, $\mathsf{P}_A$ in Alg.~\ref{alg:bipartite_algorithm}, and the first two terms in Eq.~\ref{eq:space_bipartite}. The time complexity is likewise dominated by the polynomial multiplications needed to compute the signal and ancilla polynomials, corresponding to the first term in Eq.~\ref{eq:time_bipartite}. For the symmetric case with $m_S = m_A = m/2$, $n_S = n_A = n/2$, and standard input states for which $|\mathcal{S}| = \binom{n}{n/2}$, the complexities simplify to:
\begin{equation}
	\mathrm{Space} = O\left(\left(\frac{n}{2}+1\right)^{m/2}\right) \,,
\end{equation}
\begin{equation}
	\mathrm{Time} = O\left(\frac{m^2 \, 2^n}{\sqrt{n}} \left(\frac{n}{2}+1\right)^{m/2} \log n\right) \,.
\end{equation}
The dominant contributor to both space and time complexities is that our FFT-based algorithms operate on dense tensor representations of the polynomials. Due to photon-number conservation (equivalently, polynomial homogeneity), most tensor entries are zero, making this representation highly inefficient. While sparse polynomial multiplication algorithms and GPU-parallelized methods exist~\cite{Johnson-sparse-polynomial-arithmetic, Asadi-Brandt-sparse-polynomial-arithmetic, Popescu-Garcia-multivariate-polynomial-multiplication-GPU}, we are not aware of any polynomial multiplication algorithm that simultaneously (1) utilizes sparse representations, (2) supports GPU parallelization, and (3) provides native automatic differentiation. Were such an algorithm to be developed, the efficiency of our algorithms could be significantly improved.

\begin{figure}[!htbp]
\includegraphics{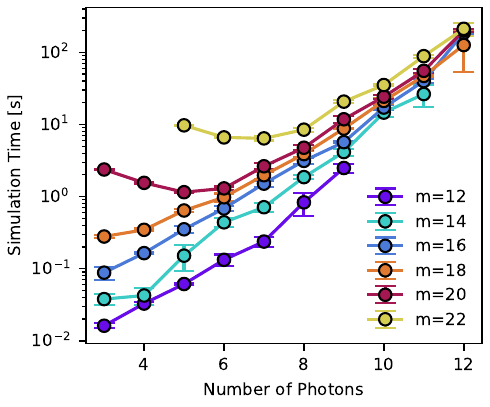}
    \caption{
    \label{fig:algorithm_scaling_vs_photons}    
    Strong simulation time. The time to compute the output state from an initial Fock state is shown as a function of the number of photons, $n$. The modes and photons were evenly divided between the signal and ancilla systems, i.e., $n_S = n_A = n/2$ and $m_S = m_A = m/2$. A cut-off dimension value of $d_{\text{cutoff}} = 3$ is used, which was the value used in our optimizations for the results presented in the main paper. Some data points are missing where the cutoff eliminated all ancilla states. The input state was taken to be $\kket{11\ldots 1 0 0 \ldots 0}$. Error bars represent fluctuations in run-times, and correspond to a single standard deviation, computed across 10 samples.
    } 
\end{figure}

To gauge practical performance, we benchmark empirical runtimes. Figure \ref{fig:algorithm_scaling_vs_photons} plots simulation time versus photon number with modes and photons split evenly across rails ($n_S=n_A=n/2$, $m_S=m_A=m/2$). We apply a per-mode cutoff $d_{\text{cutoff}}=3$---limiting occupancy to speed computation at the cost of small amplitude errors---so the measurements deviate from the pure asymptotic analysis above. The curves are roughly linear on a semi-log scale, indicating exponential growth in $n$; a simple fit suggests each added photon increases the runtime by $2.2$--$3\times$ depending on $m$. 

\subsection{5-mode Bell state generator example}
Next, we illustrate the FFT-based multiplication algorithm for the 5-mode Bell state generator example considered in Sec.~\ref{sec:5modeexample}, first considering the full statevector simulation via \textsc{PolyMulFFT}, followed by the restriction to a single sector via \textsc{BipartitePolyMulFFT}. The circuit uses $m=5$ modes and $n=4$ photons. The initial state is the Fock state $\kket{11110}$, so that the initial mode occupation numbers are ${n_{1} = \ldots = n_4 = 1}$ and $n_5 = 0$. In this case, the normalization factor $\sqrt{n_1! \ldots n_5!} = 1$ and thus the polynomial coefficients exactly match the Fock state coefficients, $T_{\bm{n}} = c_{\bm{n}}$, cf. Eq.~\ref{eq:coefficient_relation}. The polynomial corresponding to the output state is
\begin{widetext}
\begin{align}
	P(\bm{a}^{\dagger}) &= 
	\left( \sum_{i_1 = 1}^5 a_{i_1}^{\dagger} U_{i_1, 1}\right) 
	\left( \sum_{i_2 = 1}^5 a_{i_2}^{\dagger} U_{i_2, 2}\right) 
	\left( \sum_{i_3 = 1}^5 a_{i_3}^{\dagger} U_{i_3, 3}\right) 
	\left( \sum_{i_4 = 1}^5 a_{i_4}^{\dagger} U_{i_4, 4}\right) \\
	&= \frac{1}{36} 
	\left( \sqrt{2} a_1^{\dagger} + \omega^{18} a_2^{\dagger} + \sqrt{2} \omega^{18} a_3^{\dagger} + \omega^6 a_5^{\dagger} \right) 
	\left( \sqrt{2} \omega^6 a_1^{\dagger} + \omega^{16} a_2^{\dagger} + \sqrt{2} \omega^8 a_3^{\dagger} + \omega^4 a_5^{\dagger} \right)  \nonumber  \\
	& \qquad \qquad \times \left( \sqrt{2} a_1^{\dagger} + \omega^2 a_2^{\dagger} + \sqrt{2} \omega^{10} a_3^{\dagger} + \omega^{14} a_5^{\dagger} \right)
	\left( \omega^6 a_2^{\dagger} + 2 a_4^{\dagger} + \omega^6 a_5^{\dagger} \right) \nonumber \,,
\end{align}
\end{widetext}
where the explicit unitary (Eq.~\ref{eq:BSGmatrix}) has been used (and we have again reverted to our earlier convention that creation operators refer to the output modes, unless otherwise noted). When multiplied out and simplified using ${\omega = e^{i \pi / 12}}$, this can be seen to agree with Eq.~\ref{eq:BSGexample}. 

Aside from the numerical prefactor of $1/36$, each term is a degree-1 polynomial in the creation operators and may therefore be computed via \textsc{PolyMulFFT}. Each degree-1 polynomial term may be represented by a tensor; we denote these by $T^{(i)}$ for $i=1, ..., 4$. These are all zero-padded to support a maximum degree of 4, and thus have shape $(5, 5, 5, 5, 5)$. To give an explicit example, 
\begin{widetext}
\begin{equation}
	T^{(1)}_{\bm{n}} = \frac{1}{\sqrt{6}} \left( \sqrt{2} \mathbbm{1}_{\{\bm{n}=(1,0,0,0,0)\}}  + \omega^{18} \mathbbm{1}_{\{\bm{n}=(0,1,0,0,0)\}}+ \sqrt{2} \omega^{18} \mathbbm{1}_{\{\bm{n}=(0,0,1,0,0)\}} + \omega^{6} \mathbbm{1}_{\{\bm{n}=(0,0,0,0,1)\}}  \right) \,,
\end{equation}
\end{widetext}
where $\mathbbm{1}_{\{ \cdot \}}$ is the Iverson bracket. Note that these tensors are sparse -- only 3 or 4 of the $5^5 = 3125$ elements are non-zero. The resulting degree-4 polynomial may be computed by multiplying the FFT of each tensor elementwise and then taking an inverse FFT of the result, as detailed in Eq.~\ref{eq:PolyMulFFT} in the main text or in Algorithm~\ref{alg:polymulfft}.

Next, we consider the restriction to the sector containing the target Bell state, which has exactly 2 photons in the signal modes and 2 photons in the ancilla modes, $n_S = 2$. Setting $m = 5$ and using the fact that $n_1, \ldots, n_4 = 1, n_5 = 0$, Eq.~\ref{eq:sector_sum_derivation} simplifies to
\begin{equation}
  P^{(2,2)}\big( \bm{a}^\dagger \big) = \sum_{\substack{s_1+s_2+s_3+s_4=2\\ s_i\in\{0,1\}}}
  \;\prod_{i=1}^{4}\!\big(a_{\text{in}, S,i}^\dagger\big)^{s_i}
  \;\prod_{i=1}^{4}\!\big(a_{\text{in}, A,i}^\dagger\big)^{\,1-s_i}.
\end{equation}
For each term in the outer sum, a degree-2 polynomial is separately computed for each mode type (signal or ancilla) using the FFT-based algorithm described above. This can be verified by direct calculation to confirm that it agrees with the expression worked out in the main text, namely 
\begin{equation}
	P^{(2,2)}\big( \bm{a}^\dagger \big) = \frac{1}{6} \left( a_1^{\dagger} a_3^{\dagger} - a_2^{\dagger} a_4^{\dagger} \right) \left( a_5^{\dagger} \right)^2  \,,
\end{equation}
which corresponds to the desired Bell state. This is the term that is computed by \textsc{BipartitePolyMulFFT}. 

\subsection{Comparison to SLOS}
Lastly, we compare our FFT-based algorithms to the Strong Linear Optical Simulator, \textsc{SLOS}, introduced in Ref.~\cite{heurtel_strong_2023} and implemented as part of the \textit{Perceval} package~\cite{heurtel_perceval_2023}. \textsc{SLOS} comprises multiple variants---one computing the full output statevector for a given input (\textsc{SLOS\_full}), a generalized variant for arbitrary sets of inputs and outputs (\textsc{SLOS\_gen}), and a hybrid weak/strong simulation variant. For simplicity, we focus our comparison on \textsc{SLOS\_full}.

For \textsc{SLOS\_full}, the space complexity is $O\left( \left| \Phi_{m,n} \right| \right)$ and time complexity is $O \left( n \left| \Phi_{m,n} \right| \right)$, where $\left| \Phi_{m,n} \right| = \binom{n + m - 1}{m - 1}$ is the dimension of the $n$-photon Fock space over $m$ modes. In contrast, our partition-based implementation has space and time complexities given by Eqs.~\ref{eq:space_bipartite} and ~\ref{eq:time_bipartite}, respectively. For the symmetric bipartition case ($m_S = m_A = m/2$, $n_S = n_A = n/2$), these become
\begin{equation}
	\mathrm{Space} = O\left(\left(\frac{n}{2}+1\right)^{m/2}\right) \,,
\end{equation}
\begin{equation}
	\mathrm{Time} = O\left(\frac{m^2 \, 2^n}{\sqrt{n}} \left(\frac{n}{2}+1\right)^{m/2} \log n\right) \,.
\end{equation}
\textsc{SLOS\_full} achieves substantially better asymptotic scaling. The space efficiency is asymptotically optimal -- the space complexity scales linearly with the number of terms in a generic output state. 
Assuming $m$ scales with $n$---for instance, when $m = n/2$---\textsc{SLOS\_full} scales as $O(c^n \sqrt{n})$ for some constant $c \approx 2.6$, whereas our approach scales as $O(2^{3n/4} n^{n/4 + 3/2} \log n)$, with the super-exponential $n^{n/4}$ factor dominating the complexity.

However, there is a fundamental architectural difference between these approaches that makes direct performance comparison incomplete. \textsc{SLOS} iteratively builds up states through explicit loops over discrete Fock space configurations with state-dependent branching, making it not designed for automatic differentiation or GPU acceleration. In contrast, our polynomial-based representation is inherently differentiable---the circuit action is expressed as a composition of polynomial multiplications, which are naturally compatible with automatic differentiation frameworks. This enables efficient gradient computation via backpropagation, which is essential for the gradient-based optimization at the core of our automated circuit discovery pipeline. The dominant contributor to the asymptotic scaling of our FFT-based algorithms is the use of dense tensor representations, necessitated by the current absence (to the best of our knowledge) of sparse polynomial multiplication algorithms that simultaneously support GPU acceleration and automatic differentiation.

Therefore, these tools serve complementary roles: \textsc{SLOS} excels at forward simulation of fixed circuits, while our differentiable approach enables gradient-based optimization--a capability essential for automated circuit design that justifies the exponential computational overhead.

\section{Linear photonic circuits \label{app:opticalconventions}}
Here, we present our conventions for linear photonic circuits and describe a simple circuit compilation procedure that maps circuits to a canonical form in terms of non-local beamsplitters.

\subsection{Optical component conventions}
We will consider photonic circuits composed of two elementary optical components: beamsplitters (BS) and phaseshifters (P). Beamsplitters act on two modes and effectuate a $SU(2)$ transformation, whereas phaseshifters simply add a phase to a given mode. The corresponding $m$-mode transfer matrices are:
\begin{equation}
    \label{eq:beamsplitter}
    \begin{aligned}
    \mathrm{BS}(i,j; \theta, \phi_T, \phi_R) &= \mathbb{I} + 
    \begin{pmatrix}
    e^{i\phi_T} \cos\theta - 1 & -e^{-i\phi_R} \sin\theta \\
    e^{i\phi_R} \sin\theta & e^{-i\phi_T} \cos\theta - 1
    \end{pmatrix}_{(i,j)} \,, \\
    \mathrm{P}(i; \varphi) &= \mathbb{I} + (e^{i\varphi} - 1) \, \ket{i}\bra{i} \,,
    \end{aligned}
\end{equation}
The $i,j$ denote mode indices and range from $i, j \in \{0, 1, ..., m-1\}$, with $i < j$ assumed. Additionally, $\mathbb{I}$ denotes the full, $m \times m$ identity matrix.

We refer to diagonal beamsplitters (with ${\theta = 0 \pmod{\pi}}$) as \textit{trivial}, as they can be decomposed into a product of phaseshifters for each of the two modes. A second special case is $\theta = (2k+1) \pi/2$, $k \in \mathbb{Z}$, in which case the beamsplitters are purely off-diagonal and can be decomposed into a combination of a phaseshifter and a SWAP--these are referred to as \textit{SWAP-equivalent} beamsplitters. For instance,
\begin{equation}
    \mathrm{BS}(i, j; \pi/2, 0, 0) = \mathrm{P}(i; \pi) \mathrm{S}(i, j) \,,
\end{equation}
where the SWAP operation interchanges modes:
\begin{equation}
    \label{eq:swap}
    \mathrm{S}(i,j) = \mathbb{I} + (\ket{j}\bra{i} + \ket{i}\bra{j} - \ket{i}\bra{i} - \ket{j}\bra{j}) \,,
    \end{equation}

\subsection{Beamsplitter fabric}

A general photonic circuit consists of beamsplitters and phaseshifters. A \textit{beamsplitter fabric} refers to a circuit where beamsplitters act only on adjacent modes and are arranged in a regular, lattice-like pattern. Certain fabric patterns are known to be \textit{universal}, meaning they can implement arbitrary unitary transformations using only beamsplitters and phaseshifters~\cite{reck_experimental_1994, clements_optimal_2016}. Although the first stage of our pipeline utilizes a Lie-algebraic parameterization, the second stage uses such a universal fabric parameterization. Note that the first stage considers the \emph{special} unitary group $\mathrm{SU}(m)$, whereas the second stage parameterizes the broader $\mathrm{U}(m)$ group---with no loss of generality since the overall phase factor is experimentally irrelevant. 

\begin{figure*}[!htbp]
\includegraphics[width=\textwidth]{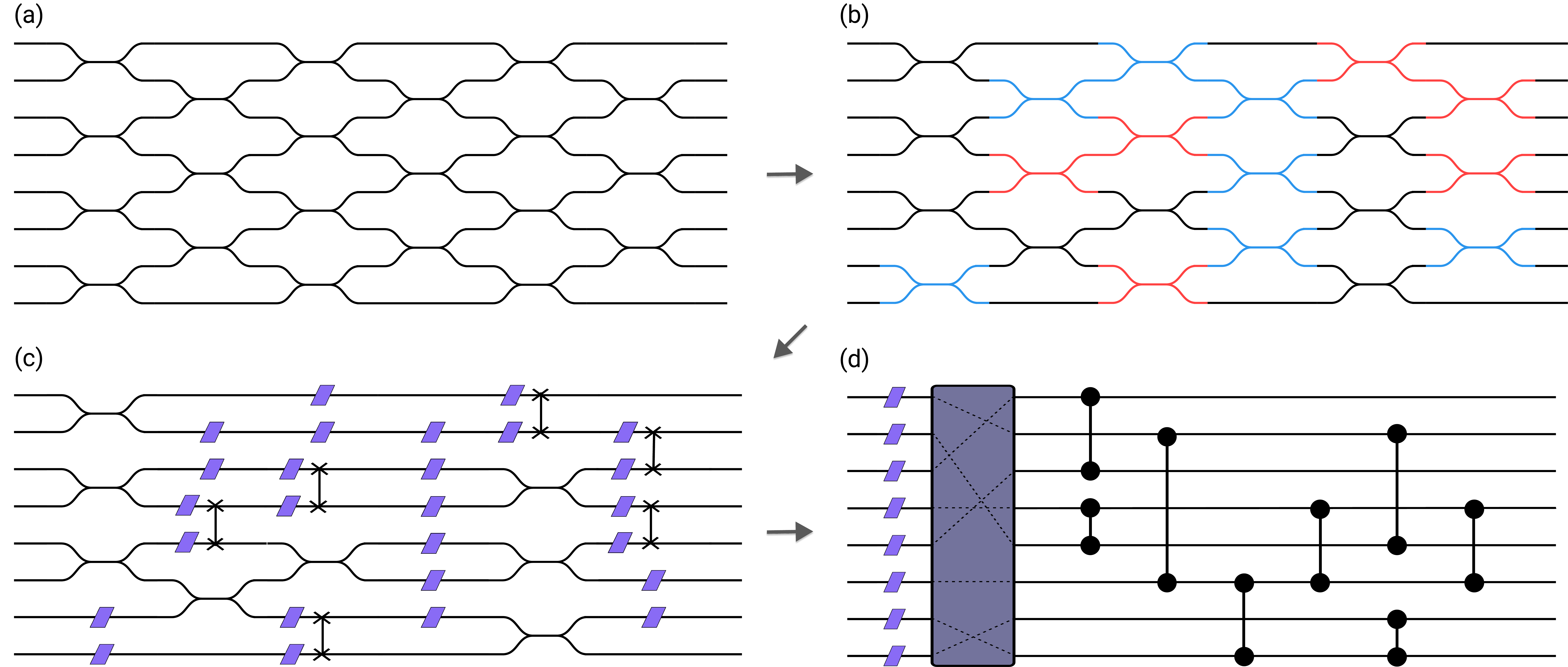}
    \caption{
    \label{fig:compilation}
    Circuit compilation: 
(i) Initial dense beamsplitter fabric obtained by applying the Clements {\em et al.} decomposition to the transfer matrix returned from Stage 1.  
(ii) A second optimization pass uses a regularized loss function to produce a sparsified fabric. Trivial beamsplitters (purely diagonal $2 \times 2$ transfer matrices) are shown in blue, and SWAP-equivalent beamsplitters (purely off-diagonal $2 \times 2$ transfer matrices) are shown in red.  
(iii) Decomposition of trivial and SWAP-equivalent beamsplitters into phaseshifters (purple quadrilaterals) and mode-SWAPs (vertical lines capped by ``x'').  
(iv) Commutation identities move all phaseshifters and SWAPs to the front of the circuit, where SWAPs combine into an overall mode permutation (gray box).  
    }
\end{figure*}

We use a variant of the Clements {\em et al.} fabric~\cite{clements_optimal_2016}, which consists of alternating layers of beamsplitters that couple adjacent modes in a brick-like pattern. Each layer $\ell$ is defined as:
\begin{equation}
    \mathcal{F}_{\ell}(f_{\ell}) = 
    \begin{cases}
        \prod_{i=1}^{\lfloor m/2 \rfloor} \mathrm{BS}(2i - 1, 2i) \,,& \ell \text{ odd} \\
        \prod_{i=1}^{\lfloor (m-1)/2 \rfloor} \mathrm{BS}(2i, 2i + 1) \,,& \ell \text{ even}
    \end{cases}
\end{equation}
where odd layers couple modes $(1,2), (3,4), \ldots$ and even layers couple modes $(2,3), (4,5), \ldots$. The complete $L$-layer fabric is given by:
\begin{equation}
    \mathcal{F}(f) = \prod_{\ell=1}^L \mathcal{F}_{\ell} (f_{\ell}) \prod_{i=1}^m \mathrm{P}(i; \varphi_i) \,,
\end{equation}
where $f_{\ell}$ denotes the collection of beamsplitter parameters in layer $\ell$, and $f$ denotes the total set of fabric parameters---the beamsplitter parameters for all layers, as well as the $m$ phaseshifter angles $\varphi_i$. We use $L = m$ layers in total.

Our implementation differs from the original Clements {\em et al.} fabric in that we use 3-parameter beamsplitters (allowing arbitrary $\mathrm{SU}(2)$ transformations) rather than their 2-parameter convention. This makes our fabric over-parameterized: while the Clements parameterization uses exactly $m^2$ real parameters to specify an $m$-mode unitary (agreeing with the dimension of the unitary group, $U(m)$), our approach uses more parameters, creating a many-to-one mapping from beamsplitter parameters to unitary matrices. Despite this redundancy, the same decomposition algorithm can be used to map any unitary matrix into our fabric parameterization.

\subsection{Photonic circuit compilation \label{app:compilation}}
While local beamsplitter fabrics are convenient for numerical optimization, it is often desirable to compile them into circuits with non-local beamsplitters that may act on arbitrary pairs of modes. This format is advantageous for both analytical interpretation and hardware implementations where connectivity is not limited to nearest neighbors.

We describe a photonic circuit compilation procedure that transforms a beamsplitter fabric into a unitarily-equivalent circuit with a canonical layered structure:
\begin{enumerate}
    \item A layer of phaseshifters, one per mode;
    \item A layer of SWAPs operations;
    \item A layer of non-local beamsplitters.
\end{enumerate}
This yields an ordered sequence of circuit instructions,
\begin{equation}
    \mathcal{C} = [\mathcal{P}, \mathcal{S}, \mathcal{BS}] \,,
\end{equation}
where $\mathcal{BS}$ contains only nontrivial and SWAP-inequivalent beamsplitters, meaning that beamsplitter angle is restricted: $\theta \neq k \pi / 2$ for any $k \in \mathbb{Z}$.

The compilation is illustrated in Fig.~\ref{fig:pipeline}~(d), and reproduced here in Fig.~\ref{fig:compilation}. First, trivial beamsplitters (which act as identities or simple phase shifts) are replaced with two phaseshifters, and SWAP-equivalent beamsplitters are rewritten in terms of phaseshifters and an explicit SWAP operation. Then, using known commutation identities, all phaseshifters and SWAPs are pushed to the beginning of the circuit. Notably, commuting a SWAP through a beamsplitter can change the modes on which the beamsplitter acts. For example,
\begin{equation}
    \mathrm{S}(i, j)\, \mathrm{BS}(j, k) = \mathrm{BS}(i, k)\, \mathrm{S}(i, j) \,.
\end{equation}

Finally, we observe that an additional simplification is possible when the circuit is restricted to act on Fock input states, as is the case in our work. For a single Fock state (as opposed to a superposition of such states), the initial layer of phaseshifters contributes only an irrelevant global phase and may therefore be omitted. Furthermore, the SWAP layer implements a fixed permutation of the modes, which can be absorbed into the input by relabeling the modes of the Fock state accordingly. The resulting compiled circuit consists solely of the nontrivial, non-SWAP-equivalent beamsplitters, acting on a possibly permuted Fock input state.

\section{Equivalence classes of stabilizer states}

In this section, we provide further details about the equivalence classes of stabilizer states considered in the main text. Denoting the set of $n$-qubit such states as $\text{Stab}_n$, the number of physically distinct such states is \cite{gross_hudsons_2006}
\begin{equation}
    \left| \text{Stab}_n \right| = 2^n \prod_{i=1}^n (2^i + 1) \,.
\end{equation}
This evaluates to $1080$ ($n=3$), $36,720$ ($n=4$), and $2,423,520$ ($n=5$). 

Consider the action of a general group $H$ of stabilizer-preserving unitaries (we reserve the symbol $G$ to refer to graphs). The group action partitions the stabilizer states into equivalence classes. Each equivalence class can be generated by constructing the orbit of a representative element under the group action: ${[\psi] = \{ h \cdot |\psi\rangle : h \in H\}}$. The stabilizer group is the set of elements of $H$ that leave a particular state invariant (up to an irrelevant global phase): ${\text{Stab}_H(|\psi\rangle) = \{h \in H : h \cdot |\psi\rangle = e^{i\phi} |\psi\rangle\}}$. By the orbit-stabilizer theorem, the size of each orbit is ${\left| [\psi] \right| = |H| / |\text{Stab}_H(|\psi\rangle)|}$. Since the orbits form a partition of $\text{Stab}_n$, we have
\begin{equation}
    \label{eq:StabECpartition}
    |\text{Stab}_n| = \sum_{k=1}^{K} |[\psi_k]| = \sum_{k=1}^{K} \frac{|H|}{|\text{Stab}_H(|\psi_k\rangle)|} \,,
\end{equation}
where $\{[\psi_k]\}_{k=1}^K$ are the distinct equivalence classes and $|\psi_k\rangle$ is a representative from each class.

The first equivalence relation we consider is the action of the $n$-qubit Pauli group, containing $4^n$ elements. In this case, the stabilizer group for any stabilizer state $|\psi\rangle$ has order $2^n$ (by definition, stabilizer states are stabilized by $2^n$ Pauli operators), meaning that the orbit size is $4^n / 2^n = 2^n$. Notably, this is independent of the specific state---all equivalence classes under Pauli operations have equal size. The numerically discovered circuits presented in Sec.~\ref{sec:graphstateresults} used this equivalence class for the set of target states in the optimization, denoted $\mathcal{T}$.

Next, we consider the case where $H$ is the local single-qubit Clifford group, $\text{Cliff}_1^{\otimes n}$ for $n$-qubit states. Since $\text{Cliff}_1$ contains 24 elements, $\text{Cliff}_1^{\otimes n}$ contains $24^n$ elements. Unlike the Pauli case, equivalence classes under local Clifford operations can have different sizes due to varying stabilizer group orders. An important result is that each equivalence class has a (generally non-unique) graph state representative, since every stabilizer state is local Clifford equivalent to some graph state. Graph states related via local complementation operations, described in Fig.~\ref{fig:localcomplementation}, correspond to the same local Clifford equivalence class. As an important distinction, the ``LC Class'' terminology in Table~\ref{tab:discovered_circuits} refers to equivalence classes of \textit{graphs} under local complementation (which maps graphs to graphs), whereas the equivalence classes discussed here are of \textit{stabilizer states}, related by the action $\text{Cliff}_1^{\otimes n}$.

As a simple check, we can use Eq.~\ref{eq:StabECpartition} and  computational enumeration to verify the consistency of the orbit sizes in the $\text{Cliff}_1^{\otimes n}$ column of Table~\ref{tab:discovered_circuits}. For simplicity, consider the case of $n=3$. There are two connected, non-isomorphic graphs: the line and triangle graphs. Both belong to the same equivalence class, as the graphs can be related by local complementation. Through brute-force enumeration, we find this equivalence class contains 432 stabilizer states. The remaining equivalence classes have disconnected graph state representatives: the empty graph consisting of 3 isolated vertices (orbit size 216), and graphs consisting of a single edge (orbit size 144 each, with 3 such graphs contributing $3 \times 144 = 432$ states total). Summing all contributions gives $432 + 216 + 432 = 1080$, which matches the total number of 3-qubit stabilizer states. By the orbit-stabilizer theorem, these orbit sizes correspond to stabilizer groups of orders $24^3/432 = 32$, $24^3/216 = 64$, and $24^3/144 = 96$, respectively. 

Related to the above check, note that when considering equivalence classes under $\text{Cliff}_1^{\otimes n}$, graph isomorphism and stabilizer state equivalence are distinct: isomorphic graphs with different vertex labelings correspond to different stabilizer states that may or may not be locally Clifford equivalent. Since $\text{Cliff}_1^{\otimes n}$ contains only local operations (no qubit permutations), the vertex labeling matters physically---requiring enumeration over labeled graphs rather than just graph isomorphism classes. This is in contrast to the local complementation equivalence classes of graphs, where one need only consider non-isomorphic graphs.

\bibliography{refs}

\end{document}

%% file: preamble.tex
\usepackage{graphicx}%
\usepackage{dcolumn}%
\usepackage{bm}%
\usepackage[T1]{fontenc}    %
\usepackage{url}            %
\usepackage{booktabs}       %
\usepackage{amsfonts}       %
\usepackage{microtype}      %
\usepackage{amssymb,amsmath,amsthm,graphicx}
\usepackage{xcolor}
\usepackage{comment}
\usepackage{multirow}
\usepackage{tabularx}
\usepackage{algorithm}
\usepackage{algpseudocode}
\usepackage{booktabs}
\usepackage{enumitem} 
\usepackage{lipsum}
\usepackage{physics}
\usepackage{bbm}
\usepackage[export]{adjustbox}
\usepackage{placeins}
\usepackage{makecell}
\usepackage[draft]{hyperref}

\newcommand{\kket}[1]{\left|\left. #1 \right\rangle\!\right\rangle}

\newcommand{\bbrakket}[2]{\left\langle\!\left\langle #1 \middle| #2 \right\rangle\!\right\rangle}

\algrenewcommand\algorithmicrequire{\textbf{Input:}}
\algrenewcommand\algorithmicensure{\textbf{Output:}}